\documentclass[11pt]{article}
\usepackage[utf8]{inputenc}
\usepackage{bbm}
\usepackage{amsmath}
\usepackage{csquotes}
\usepackage{mathtools}
\usepackage{amsthm}
\usepackage{amssymb}
\usepackage{pdflscape}
\usepackage{float}
\usepackage{units}
\usepackage{amsfonts}
\usepackage[english]{babel}
\usepackage{xfrac}
\usepackage[hang]{footmisc}
\usepackage{lipsum}
\usepackage{enumitem}
\usepackage{hyperref}
\hypersetup{colorlinks=true,allcolors=blue}
\usepackage{multirow}
\usepackage{graphicx,xcolor,textpos}
\usepackage{natbib}
\usepackage{csquotes}
\usepackage{emptypage}
\usepackage[toc,page]{appendix}
\usepackage[toc,acronym,nonumberlist]{glossaries}
\usepackage{diagbox}
\usepackage{caption}
\usepackage{subcaption}
\usepackage{float}
\usepackage{listings}
\usepackage{accents}

\DeclareMathOperator{\EX}{\mathbb{E}}

\newcommand{\diff}{\mathop{}\!d}

\newcommand{\indep}{\perp \!\!\! \perp}

\newtheorem{theorem}{Theorem}
\newtheorem*{theorem*}{Theorem (Abadie, 2003)}
\newtheorem*{theorem**}{Theorem (Chen et al., 2003)}
\newtheorem{lemma}{Lemma}

\begin{document}

\title{Estimation of the complier causal hazard ratio under dependent censoring}

\author{$\text{Gilles Crommen}^1$ \\ \url{gilles.crommen@kuleuven.be} \and $\text{Jad Beyhum}^{2}$ \\ \url{jad.beyhum@kuleuven.be} \and $\text{Ingrid Van Keilegom}^1$ \\ \url{ingrid.vankeilegom@kuleuven.be}}

\date{$^1$ORSTAT, KU Leuven, Naamsestraat 69, 3000, Leuven, Belgium \\ $^2$Department of Economics, KU Leuven, Naamsestraat 69, 3000, Leuven, Belgium}

\maketitle

\textbf{Acknowledgments:} The authors thank the National Cancer Institute for access to NCI's data collected by the HIP Breast Cancer Screening Trial (HIPB). The computational resources and services used in this work were provided by the VSC (Flemish Supercomputer Center), funded by the Research Foundation Flanders (FWO) and the Flemish Government department EWI. Moreover, the authors thank Ilias Willems and Sofia Guglielmini for their advice on high performance computing.

\bigskip \textbf{Funding:} G. Crommen is funded by a PhD fellowship from the Research Foundation - Flanders (grant number 11PKA24N). J. Beyhum acknowledges support from the FWO (Research Foundation Flanders) through the project G018725N. I. Van Keilegom acknowledges funding from the FWO and F.R.S. - FNRS (Excellence of Science programme, project ASTeRISK, grant no. 40007517), and from the FWO (senior research projects fundamental research, grant no. G047524N).

\begin{abstract}
In this work, we are interested in studying the causal effect of an endogenous binary treatment on a dependently censored duration outcome. By dependent censoring, it is meant that the duration time ($T$) and right censoring time ($C$) are not statistically independent of each other, even after conditioning on the measured covariates. The endogeneity issue is handled by making use of a binary instrumental variable for the treatment. To deal with the dependent censoring problem, it is assumed that on the stratum of compliers: (i) $T$ follows a semiparametric proportional hazards model; (ii) $C$ follows a fully parametric model; and (iii) the relation between $T$ and $C$ is modeled by a parametric copula, such that the association parameter can be left unspecified. In this framework, the treatment effect of interest is the complier causal hazard ratio (CCHR). We devise an estimation procedure that is based on a weighted maximum likelihood approach, where the weights are the probabilities of an observation coming from a complier. The weights are estimated non-parametrically in a first stage, followed by the estimation of the CCHR. Novel conditions under which the model is identifiable are given, a two-step estimation procedure is proposed and some important asymptotic properties are established. Simulations are used to assess the validity and finite-sample performance of the estimation procedure. Finally, we apply the approach to estimate the CCHR of both job training programs on unemployment duration and periodic screening examinations on time until death from breast cancer. The data come from the National Job Training Partnership Act study and the Health Insurance Plan of Greater New York experiment respectively.
\end{abstract}

\section{Introduction}

When estimating the causal effect of a binary treatment variable 
$Z$ on a right-censored duration outcome $T$, the presence of unobserved heterogeneity poses a significant challenge. Unobserved heterogeneity refers to unmeasured confounders that influence both the treatment $Z$ and the duration outcome $T$, such that $Z$ is endogenous. Consequently, the causal effect of $Z$ on $T$ cannot be identified from the conditional distribution of $T$ given $(Z,X^{\top})$, where $X$ represents a vector of observed exogenous covariates. To address the issue of endogeneity, instrumental variable (IV) methods are commonly used \citep{LATEimbensangrist1994, angrist1995two, angrist1996identification, ABADIE2003231}. These methods use external sources of variation (instruments) that influence the treatment $Z$, but are independent of the unobserved confounders affecting both $Z$ and $T$. However, in the context of censored duration outcomes, most IV approaches assume that the censoring time $C$ is (conditionally) independent of the survival time $T$. This assumption simplifies the analysis but is often unrealistic in practice and leads to biased estimates \citep{moeschberger1984consequences, emura2016gene}. In their seminal work, \citet{kaplan1958nonparametric} even stated that: ``In either case it is usually assumed in this paper that the lifetime (age at death) is independent of the potential loss time; in practice, this assumption deserves careful scrutiny". Therefore, the main objective of this paper is to propose an IV method that can identify the causal effect of an endogenous binary treatment variable on a dependently censored duration outcome.

Our motivating example is the National Job Training Partnership Act (JTPA) study, which was designed to evaluate the efficacy of publicly funded job training programs on a range of outcomes, including unemployment duration and earnings. The unemployment duration data have been analyzed under the independent censoring assumption by \citet{frandsen2015treatment} and \citet{beyhum2024instrumental} among others. In the JTPA study, individuals were randomly assigned to either a treatment group, which was eligible for job training services, or a control group, which was ineligible for 18 months. It is important to note that participants were not obliged to adhere to the treatment that they were assigned, allowing for some movement between groups. The issue of unobserved heterogeneity arises because participants may switch between the treatment and control groups in a way that is related to their time until employment but not measured (e.g. motivation to find employment). In this context, the randomized treatment assignment serves as a natural IV. The assumption of (conditional) independent censoring would be violated in cases where a participant's decision to drop out of the study is influenced by their employment status. For the stratum of fathers who reported having no job at the time of randomization, \citet{crommen2024instrumental} found a significant negative dependence between the unemployment duration and censoring time, conditional on the measured covariates. Moreover, \citet{frandsen2019testing} found that the independent censoring assumption is often questionable for unemployment duration data. Therefore, it seems likely that the assumption of (conditional) independent censoring is not satisfied for these data.

\subsection{Related literature}\label{litrelsec} 

Instrumental variable methods have been developed and applied across econometrics and related fields to address the issue of endogeneity. Seminal contributions by \citet{robins1989analysis}, \citet{manski1990nonparametric} and \citet{balke1997bounds} demonstrate that under the standard assumptions of random assignment, exclusion restrictions and monotonicity, the average treatment effect (ATE) is not point-identified. Therefore, \citet{LATEimbensangrist1994} and \citet{angrist1996identification} proposed focusing on specific subpopulations rather than attempting to identify effects for the entire population. In particular, they demonstrated that the IV framework could identify the local average treatment effect (LATE), also referred to as the complier average causal effect, for a subgroup of individuals called the compliers. The identification strategy is based on a monotonicity assumption, which rules out the existence of another subgroup called the defiers. Compliers are defined as those whose treatment status is directly influenced by the instrument, making the treatment assignment effectively act as a randomized experiment within this subgroup. Subsequent work by \citet{ABADIE2003231} introduced a new class of IV estimators designed for linear and nonlinear treatment response models with covariates, providing additional flexibility. To estimate population-level causal effects, researchers can impose additional parametric assumptions \citep{efron1991compliance} or replace the monotonicity assumption with the rank invariance assumption as introduced by \citet{chernozhukov2005iv}.

Most instrumental variable methods for right-censored duration outcomes assume that the duration time $T$ is (conditionally) independent of the censoring time $C$. This assumption is known as independent censoring, or conditional independent censoring if it is assumed conditional on the measured covariates. Some non-parametric approaches are given by \citet{frandsen2015treatment}, \citet{SantAnnaPedroH.C2016PEwR} and \citet{BeyhumJad2021NIRW}, while other approaches such as   \citet{BijwaardGovertE2005Cfsc}, \citet{TchetgenTchetgenEricJ2015IVEi}, \citet{ChernozhukovVictor2015Qrwc}, \citet{LiJialiang2015Ivah}, \citet{kianian2021causal}, \citet{beyhum2024instrumental} and \citet{tedesco2023instrumentalvariableestimationproportional} are semiparametric. Moreover, \citet{van2020nonparametric} and \citet{beyhum2024instrumentaldynamic} consider the problem of identifying and estimating the causal effect of a dynamic treatment. The problem of endogeneity has also been discussed in a competing risks framework by \citet{RichardsonAmy2017Nbiv}, \citet{ZhengCheng2017Ivwc}, \citet{martinussen2020instrumental} and \citet{beyhum2023nonparametric} among others.  

Since the (conditional) independent censoring assumption is often unrealistic and empirically untestable, models that allow for dependent censoring have been developed. Among the most popular approaches are those based on copulas, where a first line of research focuses on fully known copulas \citep{ZHENGMING1995Eoms, Rivest2001AMA, BraekersRoel2005Acef, HuangXuelin2008RSAw, SujicaAleksandar2018Tcei}. A copula being fully known means that the association parameter specifying the dependence between $T$ and $C$ is known. Since the assumption of a fully known copula is usually equally unrealistic in practice as the independent censoring assumption, \citet{czado2021dependent} introduced a new approach that does not require the association parameter of the copula to be specified. However, this approach requires the marginals for $T$ and $C$ to be fully parametric to identify the association parameter. This approach was extended in various ways by \citet{DeresaNegeraWakgari2020Fpmf}, \citet{DERESA2020106879diftypecens} and \citet{semiparderesa2020} among others. More recently, \citet{deresa2021copulacox} showed that (under some additional assumptions regarding the copula function and the covariates) the model remains identified when the conditional marginal distribution of $T$ follows a semi-parametric proportional hazards model. The additional assumption on the covariates requires that, given 2 distinct continuous covariates $X_1$ and $X_2$, the conditional distribution of $T$ does not depend on $X_1$ and the conditional distribution of $C$ does not depend on $X_2$\footnote{Note that this assumption is not explicitly mentioned by \citet{deresa2021copulacox}, but is necessary for their identifiability proof.}. This type of covariate restriction was also used by \citet{deresa2024semiparametric} to allow for both $T$ and $C$ to follow a semi-parametric transformation model, and by \citet{hiabu2025identifiability} to allow for non-parametric marginals. A non-parametric approach that does not rely on this type of covariate restriction has been proposed by \citet{Lo_Wilke_2017}, but they only identify the sign of covariate effects on the marginals. Note that all of these approaches assume that the treatment is exogenous.

The body of literature addressing instrumental variable methods in the context of dependent censoring remains relatively limited. \citet{KhanShakeeb2009Ioec} examined an endogenously censored regression model. However, their framework imposes a restrictive support condition (Assumption IV2, page 110) concerning the relationship between the instruments and covariates. Moreover, \citet{blanco2020bounds} investigated treatment effects on duration outcomes in the presence of censoring, selection, and noncompliance. Instead of providing point estimates, their work derives bounds for the causal effect. More recently, \citet{ying2024proximal} introduced a proximal survival analysis framework, which draws on ideas from proximal causal inference. This approach addresses dependent censoring but requires the analyst to classify observed covariates into three specific categories, which can be challenging in practice. Furthermore, \citet{crommen2024instrumental} proposed a fully parametric model that identifies the causal effect of an endogenous treatment variable on a potentially dependently censored event time. Their method employs a control function approach to handle the endogeneity issue, and the possible dependent censoring is taken into account by relying on the strong parametric assumption of bivariate normal error terms of the joint regression model for the logarithms of $T$ and $C$. To make the normality assumption more realistic, \citet{rutten2024flexible} extended this work to a competing risks framework where different power transformations are applied to each risk.

\subsection{Approach}

We propose an IV method that allows for dependent censoring, enabling the identification and estimation of causal effects in settings where the (conditional) independent censoring assumption may not hold.  More precisely, we are interested in the complier causal hazard ratio (CCHR), which is the ratio of compliers' hazard function when the treatment is received versus when it is not. As mentioned before, the compliers are a latent subgroup who adhere to their assigned treatment status. However, a major challenge is that the compliers are a latent subgroup, meaning we cannot directly identify if an individual belongs to the compliers from the observed data. This is because, for any given individual, we only observe their treatment choice under their assigned treatment group. To overcome this issue, we rely on a monotonicity assumption to leverage the weighting scheme from Theorem 3.1 by \citet{ABADIE2003231}. This theorem establishes a connection between the conditional moment of any measurable real function of the observed data for the compliers and the unconditional moment. The monotonicity assumption rules out the existence of defiers, a latent subgroup for whom the treatment choice is always different from the assigned treatment. Note that this assumption is automatically satisfied when there is only one-sided noncompliance. To account for the potential dependence between $T$ and $C$, conditional on the measured covariates, it is assumed on the stratum of compliers that $T$ follows a semiparametric proportional hazards model, $C$ follows a fully parametric model and their joint distribution is modeled by a parametric copula. In addition, we allow for an independent right censoring time $A$ (e.g. administrative censoring) such that only the minimum of $T,C$ and $A$ is observed. This approach enables us to achieve the following contributions:
\begin{enumerate}
	\item[(i)] When $Z$ is exogenous, meaning that everyone is a complier, we demonstrate that the model is identifiable under less restrictive conditions than those imposed by \citet{deresa2021copulacox}. More specifically, we remove the need for the covariate restriction discussed in Section \ref{litrelsec} by substituting it with two alternative conditions on the copula function. We verify that the copula functions meeting the identifiability requirements of \citet{deresa2021copulacox} also satisfy our proposed conditions, thereby relaxing the identifiability result by removing the untestable covariate restriction. Moreover, we verify our identifiability conditions for additional copula functions to allow for more modeling flexibility.
	\item[(ii)] When $Z$ is endogenous, we propose a two-step estimator that is shown to be consistent and asymptotically normal. Note that we do not only estimate the covariate effects and the CCHR, which quantifies the causal effect of the treatment on the compliers’ hazard function, but also the association parameter of the copula model for the compliers, which characterizes the strength of dependence between $T$ and $C$ after accounting for measured covariates. Therefore, the association parameter does not need to be specified. This also implies that conditional independent censoring is a special case of our model. This is achieved by incorporating the weighting scheme developed by \citet{ABADIE2003231} in both the estimator for the baseline cumulative hazard function and the likelihood, resulting in a weighted maximum likelihood approach. The weights are estimated non-parametrically in the first stage, after which the parameters of interest are estimated by maximizing the weighted pseudo profile-likelihood function that results from plugging in the weighted estimator for the baseline cumulative hazard function. The finite-sample performance of the estimator is investigated in various simulation settings such as different copula and marginal combinations, misspecification, a small ratio of compliers and a range of sample sizes.
	\item[(iii)] We apply our method to evaluate the causal effect of job training services on unemployment duration using data from the well known National Job Training Partnership Act (JTPA) study. Moreover, we also investigate the causal effect of breast cancer screening trials on breast cancer mortality using data from the Health Insurance Plan of Greater New York experiment.
\end{enumerate}

\subsection{Outline}

In Section \ref{sectthemodel}, we introduce the potential outcomes framework and specify the model.  The identification is discussed in Section \ref{Identification} and Section \ref{estimationch} describes the two-step estimation procedure. Section \ref{asymptotics} establishes the asymptotic properties of the proposed estimator. Simulation results and an empirical application regarding the effect of job training programs on unemployment duration are described in Sections \ref{sectsimresul} and \ref{sectdataap}, respectively. The proofs, technical details, extra simulation results and a second data application can be found in the Supplementary Material. The R code can be found at \url{https://github.com/GillesCrommen/CCHR}. 

\section{The model}\label{sectthemodel}
Let $T$ and $C$ be the duration and right censoring time respectively. We will allow for the variables $T$ and $C$ to be dependent on each other, even after conditioning on the measured covariates. The covariates that influence both $T$ and $C$ are given by $(Z,X^{\top})$, where $Z$ and $X$ are of dimension 1 and $m$ respectively. Note that $X$ does not include an intercept and that $Z$ represents an endogenous binary treatment variable for which a binary instrumental variable, denoted by $W$, exists. In addition, we allow for another right censoring time $A$, which is independent of $T,C$ given the measured covariates and the measured covariates themselves. A common example of observing $A$ is when a subject is administratively censored. Because $T, C$ and $A$ censor each other, only one of them is observed through the follow-up time $Y = \text{min}\{T, C, A\}$ and the censoring indicators $\Delta_1 = \mathbbm{1}(Y = T)$ and $\Delta_2 = \mathbbm{1}(Y = C)$, where $\mathbbm{1}(\cdot)$ is the indicator function. 

\subsection{Potential outcomes}

To state the necessary assumptions, we will use the potential outcome framework as described by \citet{rubin1974estimating}, \citet{LATEimbensangrist1994} and \citet{ABADIE2003231} among others. Firstly, let $Z_{W=w}$ denote the potential treatment selection under instrument $W=w$, where $w \in \{0,1\}$, such that the observed treatment is given by $Z=W Z_{W=1} + (1-W) Z_{W=0}$. Further, let $Y_{Z=z}$ denote the potential outcome given $Z=z$, where $z \in \{0,1\}$, such that the observed outcome is given by $Y=Z Y_{Z=1} + (1-Z) Y_{Z=0}$. Using this notation, we can classify subjects into four latent subgroups: compliers ($Z_{W=1} > Z_{W=0}$), always takers ($Z_{W=1} = Z_{W=0} = 1$), never takers ($Z_{W=1} = Z_{W=0} = 0$) and defiers ($Z_{W=1} < Z_{W=0}$). We will restrict our attention to the compliers, a subgroup for whom the treatment choice acts as randomization. For ease of notation, let $G$ denote a latent variable that indicates the subgroup (i.e. $G=g$ where $g \in \{co,at,nt,df\}$ indicates compliers, always takers, never takers and defiers respectively). We will make the following standard IV assumptions for almost all values of $X$:
\begin{enumerate}[label=(A\arabic*),resume,left=0.25\leftmargin]
    \item $( T_{Z=1}, T_{Z=0},C_{Z=1}, C_{Z=0},Z_{W=1},Z_{W=0}) \text{ are jointly independent of } W \mid X$. \label{AindepIV}
    \item $\mathbb{P}(Z_{W=0} > Z_{W=1} \mid X)  = 0$. \label{Amonotonicity}
    \item $0<\mathbb{P}(W=1 \mid X)<1$ and $\mathbb{P}(Z=1 \mid W = 1,X) > \mathbb{P}(Z=1 \mid W = 0,X)$. \label{Afirststage}
\end{enumerate}
Assumption \ref{AindepIV} states that both the potential outcomes and the potential treatments are independent of the instrument given $X$. Note that this assumption is the combination of two commonly made requirements in an IV framework. Firstly, $W$ follows a random assignment conditional on $X$ such that we can identify the causal effect of the instrument on the treatment, which is commonly known as an exogeneity assumption. Secondly, the potential outcomes are not directly affected by the instrument, which is commonly known as an exclusion restriction. Therefore, the instrument only affects the outcome through variation in the treatment selection. Note that this assumption is made on both the potential outcomes of the duration and censoring time, which is a consequence of allowing for some dependence between $T$ and $C$.  Assumption \ref{Amonotonicity} is a standard monotonicity assumption that rules out the existence of defiers. This assumption is automatically satisfied when individuals in the control group ($W=0$) cannot access the treatment. Finally, the first part of Assumption \ref{Afirststage} requires non-trivial assignment of the instrument, while the second part is a standard first-stage assumption. Note that Assumption \ref{Afirststage} implies that the fraction of compliers is non-zero. 

\subsection{Model specification}\label{modelspec}

We will further assume that:
\begin{enumerate}[label=(A\arabic*),resume,left=0.25\leftmargin]
    \item $T,C$ and $A$ are absolutely continuous random variables. \label{absolutecontinuity}
    \item $(T,C) \indep A \mid Z,X$, where $\indep$ denotes statistical independence.  Moreover, $A \indep Z,X$ and the censoring by $A$ is non-informative such that the distribution of $A$ does not depend on any of the model parameters. \label{Anoninfcens}
    \item The covariance matrix of $(Z,X^{\top})$ has full rank. \label{fullrank}
    \item There exists an $\varepsilon > 0$ such that the probabilities $\mathbb{P}(\Delta_1 =1 \mid Y = y, Z=z,X=x,G=co)$ and $\mathbb{P}(\Delta_2 = 1 \mid Y = y, Z=z,X=x,G=co)$ are strictly positive for all $(z,x^{\top})$ and for all $ y \in (0,\varepsilon)$. \label{Astrpos}
    \item\label{positiveprobassumptionwithtau} There exists a finite $\bar{\tau} \in \mathbb{R}_{>0}$ such that $\mathbb{P}(\min\{T,C\} > \bar{\tau} \mid Z=z,X=x,G=co) > 0$ for all $(z,x^{\top})$. 
\end{enumerate}
Assumptions \ref{absolutecontinuity}-\ref{fullrank} are standard assumptions in the survival analysis literature and Assumption \ref{Astrpos} is necessary for the identification of the model. Lastly, Assumption \ref{positiveprobassumptionwithtau} is required to ensure that the denominator of the estimator for the baseline cumulative hazard function in Section \ref{estimatingchf} is non-zero. Note that because of Assumption \ref{positiveprobassumptionwithtau}, the endpoint of $\mathcal{Y}$, where $\mathcal{Y} \subseteq \mathbb{R}_{> 0}$ is the support of $Y$, should be greater than $\Bar{\tau}$. It is important to emphasize that the following model specifications are only assumed for the subgroup of compliers.

We start by noticing that Assumption \ref{AindepIV} and $W=Z$ for the compliers imply that
\begin{align*}
    F_{T \mid Z,X,G}(t\mid z,x,co) =\mathbb{P}(T\leq t\mid Z=z, X=x, G=co) & = \mathbb{P}(T_{Z=z}\leq t\mid W=1, X=x, G=co) \\ &= \mathbb{P}(T_{Z=z}\leq t\mid X=x, G=co),
\end{align*}
such that, in the population of the compliers, the causal effect of $Z$ on $T$ can be identified from $F_{T \mid Z,X,G}(t\mid z,x,co)$. Following \citet{deresa2021copulacox}, a proportional hazards model (see \citet{cox1972regression} for details) is assumed for the conditional distribution of $T$ given $(Z,X^{\top})$ for the subgroup of compliers:
$$
F_{T \mid Z,X,G}(t\mid z,x,co) =1-\exp\{-\Lambda(t\mid G=co) \exp(z \alpha + x^\top \beta)\},
$$
where $\mu = (\alpha,\beta^\top) \in \mathcal{M} \subset \mathbb{R}^{m+1} $ is a vector of regression coefficients and $\Lambda(t \mid G=co)$, with $\Lambda(0 \mid G= co)=0$, is an unknown strictly increasing differentiable baseline cumulative hazard function. Further, let $\lambda(t\mid G=co)=\frac{ \diff \Lambda(t\mid G=co)}{\diff t}$ be the baseline hazard function for the compliers and 
$$
f_{T \mid Z,X,G}(t\mid z,x,co)=\lambda(t\mid G=co)\exp(z \alpha + x^\top \beta)\exp\left\{-\Lambda(t\mid G=co) \exp(z \alpha + x^\top \beta)\right\},
$$ the conditional density function of $T$ given $(Z,X^{\top})$ for the compliers.
This allows us to define the conditional hazard function for the compliers as 
$$
\lambda(t\mid Z=z,X=x,G=co) = \frac{f_{T \mid Z,X,G}(t\mid z,x,co)}{1-F_{T \mid Z,X,G}(t\mid z,x,co)}= \lambda(t\mid G=co)\exp(z \alpha+ x^\top \beta).
$$
From this it is clear that $$
\alpha =\log\lambda(t\mid Z=1,X=x,G=co)-\log\lambda(t\mid Z=0,X=x,G=co),
$$
which shows that $\exp(\alpha)$ has a causal interpretation as the ratio of compliers' hazard function when the treatment is received versus when the treatment is not received. This means that when $\alpha$ is significantly greater (smaller) than zero, the treatment reduces (increases) the survival time of the compliers. We will refer to the causal effect of interest, $\exp(\alpha)$, as the complier causal hazard ratio (CCHR).

For the conditional distribution of the censoring time $C$ of the compliers, given $(Z,X^{\top})$, a fully parametric model is assumed:
$$ F_{C \mid Z,X,G}( \cdot \mid z,x,co) \in \big\{ F_{C \mid Z,X,G}( \cdot \mid z,x,co ; \eta) : \eta \in \mathcal{H} \big\},$$
for a finite-dimensional parameter space $\mathcal{H}$, where $F_{C \mid Z,X,G}(c\mid z,x,co)=\mathbb{P}(C\leq c\mid Z=z,X=x,G=co)$ and $f_{C \mid Z,X,G}(c\mid z,x,co)$ the corresponding conditional density for the compliers. This parametric model will be necessary to ensure the identifiability of the model. To account for the possible dependence between $T$ and $C$, given $(Z,X^{\top})$ for the compliers, we propose to use a bivariate copula $\mathcal{C}$. The joint distribution function of $T$ and $C$, conditional on $(Z,X^{\top})$ for the compliers, can therefore be modeled as
$$
\mathbb{P}(T \leq t, C\leq c\mid Z=z,X=x,G=co)=\mathcal{C}\left(F_{T \mid Z,X,G}(t\mid z,x,co),F_{C \mid Z,X,G}(c\mid z,x,co)\right),
$$
with $t,c > 0$ and for some copula $\mathcal{C}$. The copula $\mathcal{C}$ is defined as a bivariate distribution function with uniform marginals over the unit interval. Note that Assumption \ref{absolutecontinuity} implies that the copula function is unique \citep{sklar1959fonctions}. We will assume that $\mathcal{C}$ is a parametric copula, meaning that
$$
\mathcal{C} \in \{ \mathcal{C}_\xi : \xi \in \Xi \},
$$
for some parameter space $\Xi$. Moreover, for $u,v \in [0,1]$ let 
$$
\zeta_{1}(u,v)=\frac{\partial}{\partial u}\mathcal{C}(u,v), \quad \zeta_{2}(u,v)=\frac{\partial}{\partial v}\mathcal{C}(u,v).
$$
It can be shown that (see \cite{czado2021dependent}):
\begin{align*}
& \mathbb{P}(T \leq t \mid C = c, Z=z,X=x, G=co) = \zeta_{2}\big(F_{T \mid Z,X,G}(t\mid z,x,co), F_{C \mid Z,X,G}(c\mid z,x,co)\big), \\ & \mathbb{P}(C \leq c \mid T = t, Z=z,X=x, G=co) = \zeta_{1}\big(F_{T \mid Z,X,G}(t\mid z,x,co), F_{C \mid Z,X,G}(c\mid z,x,co)\big).    
\end{align*}
The distribution of $A$, denoted by $F_A(a) = \mathbb{P}(A \leq a)$ with $f_A(a)$ the corresponding density, is left completely unspecified. 

Further, let $F_{Y,\Delta_1,\Delta_{2} \mid Z,X,G}(y,\delta_1,\delta_2 \mid z,x,co)$ denote the sub-distribution function of $(Y,\Delta_1,\Delta_2)$ given $(Z,X^{\top})$ for the compliers, that is
$$
F_{Y,\Delta_1,\Delta_{2} \mid Z,X,G}(y,\delta_1,\delta_2 \mid z,x,co) = \mathbb{P}(Y \leq y,\Delta_1 = \delta_1,\Delta_2 = \delta_2  \mid Z=z, X=x,G=co),
$$
with the corresponding sub-density $f_{Y,\Delta_1,\Delta_{2} \mid Z,X,G}(y,\delta_1,\delta_2 \mid z,x,co)$, for which it easily can be shown that:
\begin{align*}
     f_{Y,\Delta_1,\Delta_{2} \mid Z,X,G}(y,1,0 \mid z,x,co) & =\big\{1-\zeta_{1}\big( F_{T \mid Z,X,G}(y\mid z,x,co),F_{C \mid Z,X,G}(y\mid z,x,co)\big)\big\} \\ & \quad \times f_{T \mid Z,X,G}(y\mid z,x,co)\{1-F_A(y)\}, 
\end{align*}
\begin{align*}
f_{Y,\Delta_1,\Delta_{2} \mid Z,X,G}(y,0,1 \mid z,x,co) &=\big\{1-\zeta_{2}\big( F_{T \mid Z,X,G}(y\mid z,x,co),F_{C \mid Z,X,G}(y\mid z,x,co)\big)\big\} \\ & \quad \times f_{C \mid Z,X,G}(y\mid z,x,co)\{1-F_A(y)\}, 
\end{align*}
and
$$
f_{Y,\Delta_1,\Delta_{2} \mid Z,X,G}(y,0,0 \mid z,x,co)= S(y \mid z,x,co) \times f_A(y),
$$
where
\begin{align*}
S(y \mid z,x,co) & = 1- F_{T \mid Z,X,G}(y\mid z,x,co) - F_{C \mid Z,X,G}(y\mid z,x,co) \\ & \quad + \mathcal{C}\big(F_{T \mid Z,X,G}(y\mid z,x,co),F_{C \mid Z,X,G}(y\mid z,x,co)\big).    
\end{align*}
Lastly, we have that  
\begin{align*}
F_{Y \mid Z,X,G}(y \mid z,x,co) & = \mathbb{P}(Y \leq y \mid Z=z,X=x,G=co)\\ & = 1-\mathbb{P}(T > y, C > y\mid Z=z,X=x,G=co)\times \mathbb{P}(A > y)\\ & = 1-S(y \mid z,x,co)\{1-F_A(y)\}.
\end{align*}
Note that when we want to stress the dependence of these functions on their respective parameters, we will add them to their notation (e.g. $S_{\mu,\Lambda,\eta,\xi}$ instead of $S$).

\section{Identification}\label{Identification}

In this section, we will discuss the identifiability of our model. Note that we have two identifiability issues. Firstly, we cannot identify if a particular individual belongs to the subgroup of compliers since only one of $Z_{W=0}$ and $Z_{W=1}$ is observed. Secondly, we want to show that two different parameter vectors result in two different joint distributions of $S=(Y,\Delta_1,\Delta_2,Z,X^\top)^\top$ for the subgroup of compliers. 
\subsection{Identifying expectations for compliers}\label{identcomp}
We will start by restating Theorem 3.1 by \citet{ABADIE2003231}. This theorem shows that even though we cannot identify if a particular individual belongs to the subgroup of compliers, we can identify expectations for compliers.
\begin{theorem*}
   Let $g(\cdot)$ be any measurable real function of $S$ such that $\EX\lvert g(S)\rvert < \infty$. Under Assumptions \ref{AindepIV}-\ref{Afirststage}, we have that
   \begin{equation*}
   \EX\left[g(S) \mid G = co\right] = \frac{1}{\mathbbm{P}(G=co)}\EX\left[\omega \times g(S) \right],       
   \end{equation*}
   with
   $$
   \omega = 1 - \frac{Z(1-W)}{\mathbbm{P}(W=0 \mid X) }- \frac{(1-Z)W}{\mathbbm{P}(W=1 \mid X) }.
   $$  
\end{theorem*}
Because of this theorem, we have that any statistical characteristic that is defined in terms of moments of $S$ is identified for the compliers.
\subsection{Identifying the model parameters}
Let $\mathcal{S} = [0,\Bar{\tau}] \times \{0,1\}\times \{0,1\}\times \{0,1\} \times \mathcal{X}$ be the support of $S$, where $\mathcal{X} \subset \mathbb{R}^{m}$ is the support of $X$. Note that we have restricted $\mathcal{Y}$ to $[0,\Bar{\tau}]$ as this is the relevant support for $Y$. The problem of identifiability under dependent censoring has been studied by \citet{deresa2022copula} and \citet{czado2021dependent} for a fully parametric model and by \citet{deresa2021copulacox} for a semi-parametric model. To show the identifiability of the model for the subgroup of compliers, we will need the following conditions:

\begin{enumerate}[label=(C\arabic*),left=0.25\leftmargin]
\item\label{identifiabilityC} For all $\eta_1,\eta_2 \in \mathcal{H}$, we have that:
$$
\lim_{c \to 0}\frac{f_{C,\eta_1\mid Z,X,G}(c\mid z,x,co)}{f_{C,\eta_2\mid Z,X,G}(c\mid z,x,co)}= 1 \text{ for all } (z,x^{\top}) \iff \eta_1 = \eta_2. 
$$
\item\label{identifiabilitydependence1} For all $(\mu,\eta,\xi,\Lambda) \in \mathcal{M} \times \mathcal{H} \times \Xi \times \mathcal{L}$, we have that:
$$
\lim_{y \to 0} \zeta_{k,\xi}\big(F_{T \mid Z,X,G,\mu,\Lambda}(y\mid z,x,co) , F_{C \mid Z,X,G,\eta}(y\mid z,x,co)\big) = 0 \text{ for all } (z,x^{\top}) \text{ with } k=1,2.
$$
\item\label{identifiabilitydependence2} For all $(\mu,\eta,\xi_k,\Lambda_k) \in \mathcal{M} \times \mathcal{H} \times \Xi \times \mathcal{L} $, $k=1,2$, such that $\lim_{t \to 0}\frac{\lambda_1(t)}{\lambda_2(t)}=1$, we have that:
\begin{align*}
 \lim_{y \to 0}\frac{\zeta_{2,\xi_1}\big(F_{T \mid Z,X,G,\mu,\Lambda_1}(y\mid z,x,co) , F_{C \mid Z,X,G,\eta}(y\mid z,x,co)\big)}{\zeta_{2,\xi_2}\big(F_{T \mid Z,X,G,\mu,\Lambda_2}(y\mid z,x,co) , F_{C \mid Z,X,G,\eta}(y\mid z,x,co)\big)} = 1 \text{ for all } (z,x^{\top}) \iff \xi_1 = \xi_2.
\end{align*}
\item\label{identifiabilitydependence3} For all $\xi \in  \Xi$, we have that:
\begin{align*}
 c_{\xi}(u,v) > 0 \text{ for all } u,v \in (0,1),
\end{align*}
where $c_{\xi}$ is the density of the copula $\mathcal{C}_{\xi}.$
\end{enumerate}
It has already been shown by \citet{czado2021dependent} that Condition \ref{identifiabilityC} is satisfied for the parametric families of log-normal, log-logistic, log-Student-t and Weibull densities. More recently, it was shown by \citet{delhelle2024copula} that this condition is also satisfied for the parametric family of Gamma densities. Furthermore, let
the strict Clayton copula be given by
$$
\mathcal{C}_{\xi}(u,v) = (u^{-\xi} + v^{-\xi} -1)^{-1/\xi}, \quad \xi > 0,
$$ and the rotated copulas by
\begin{align*}
	& \mathcal{C}^{90}_{\xi}(u,v) = v - \mathcal{C}_{\xi}(1-u,v), \\& \mathcal{C}^{180}_{\xi}(u,v) = u + v - 1 + \mathcal{C}_{\xi}(1-u,1-v), \\& \mathcal{C}^{270}_{\xi}(u,v) = u - \mathcal{C}_{\xi}(u,1-v),
\end{align*}
where the superscript indicates the angle of rotation. For ease of notation, let Clayton($\gamma$) be the rotated strict Clayton copula with $\gamma$ the angle of rotation. We now have the following result, for which the proof can be found in Section A of the Supplementary Material.
\begin{lemma}\label{copulaslemma}
	Conditions \ref{identifiabilitydependence1}-\ref{identifiabilitydependence3} are satisfied by
	\begin{enumerate}[label=(\arabic*),left=0.25\leftmargin]
		\item the Frank, Joe, Clayton(90), Clayton(180) and Clayton(270) copulas, independently of the \\ marginal distributions;
		\item the Gumbel and Gaussian copula if $$\lim_{y \to 0}\frac{f_{T \mid Z,X,G,\mu,\Lambda}(y\mid z,x,co)}{f_{C \mid Z,X,G,\eta}(y\mid z,x,co)} \in \mathbb{R}_{>0} \text{ for all } (z,x^{\top}) \text{ and for all } (\mu,\Lambda,\eta) \in \mathcal{M}   \times \mathcal{L} \times \mathcal{H},$$
		with $\Xi \subseteq (-1,1)$ for the Gaussian copula.
	\end{enumerate}
\end{lemma}
Although the conditions for the Gumbel and Gaussian copula given in Lemma \ref{copulaslemma} seem different from the ones given in Lemma 3.1 by \citet{deresa2021copulacox}, they are equivalent. The advantage of presenting the conditions in this form is that they are much easier to interpret. Further, it is to be noted that the non-rotated Clayton copula is not mentioned by Lemma \ref{copulaslemma}. This is because Condition \ref{identifiabilitydependence1} can never hold for the Clayton copula, as it would require that $\lim_{y \to 0}F_{T \mid Z,X,G}(y\mid z,x,co)/F_{C \mid Z,X,G}(y\mid z,x,co) \text{ converges to } 0 \text{ and } \infty \text{ at the same time. }$
However, this does not necessarily mean that the model is not identified, since these conditions are only sufficient for identification. We now have the following identifiability theorem:
\begin{theorem}\label{theoremident}
    Under Assumptions \ref{absolutecontinuity}-\ref{positiveprobassumptionwithtau} and Conditions \ref{identifiabilityC}-\ref{identifiabilitydependence3}, suppose that, for the subgroup of compliers, the pair $(T_k,C_k,A_k)$, $k=1,2$, satisfies the model described in Section \ref{modelspec} with parameter vector $(\mu_k,\eta_k,\xi_k,\Lambda_k) \in \mathcal{M} \times \mathcal{H} \times \Xi \times \mathcal{L}$ and suppose that $(Y_k,\Delta_{1,k},\Delta_{2,k})$ given $(Z,X^\top)$ have the same distribution for the subgroup of compliers, then
    $$
    \mu_1 = \mu_2,\eta_1=\eta_2,\xi_1=\xi_2,\Lambda_1=\Lambda_2.
    $$
\end{theorem}
The proof of Theorem \ref{theoremident} can be found in Section A of the Supplementary Material. It is to be noted that Conditions \ref{identifiabilityC}-\ref{identifiabilitydependence3} are similar to the ones given by \citet{deresa2021copulacox}, but differ as Condition \ref{identifiabilitydependence2} is about the quotient of $\zeta_{2,\xi_1}$ and $\zeta_{2,\xi_2}$, instead of the respective copula densities, and we have added Condition \ref{identifiabilitydependence3}. By using these conditions, we avoid having to assume that, given 2 distinct continuous covariates $X_1$ and $X_2$, the conditional distribution of $T$ does not depend on $X_1$ and the conditional distribution of $C$ does not depend on $X_2$. This covariate restriction is not explicitly mentioned by \citet{deresa2021copulacox}, but is necessary for their identification proof. We have verified in Lemma \ref{copulaslemma} that the copula functions meeting the identifiability requirements of \citet{deresa2021copulacox} also satisfy our proposed conditions, thereby relaxing the identifiability result by removing the untestable covariate restriction. Moreover, Lemma \ref{copulaslemma} verifies the identifiability conditions for additional copula functions to allow for more modeling flexibility.

\section{Estimation}\label{estimationch}
The observed data consists of $n$ i.i.d. realizations of $O = (Y,\Delta_1,\Delta_2,Z,X^\top,W)$, which we denote by $O_i = \{Y_i,\Delta_{1,i},\Delta_{2,i},Z_i,X_i^\top,W_i\}_{i=1,...,n}$. If we let $\theta = (\mu,\eta,\xi)$, the profile pseudo-likelihood function for the compliers can be given by 
\begin{align*}
 L(\theta,\Lambda) & =  \prod^n_{i=1}f_{Y,\Delta_1,\Delta_{2} \mid Z,X,G,\theta,\Lambda}(Y_i,\Delta_{1,i},\Delta_{2,i} \mid Z_i,X_i,co) \\ & \propto  \prod^n_{i=1} \Big[S_{\theta,\Lambda}(Y_i \mid Z_i,X_i, co) \Big]^{(1-\Delta_{1,i})(1-\Delta_{2,i})} \\ & \quad \times \Big[\lambda(Y_i \mid G=co)\exp( Z_i \alpha+X_i^\top \beta)\exp\big\{-\Lambda(Y_i \mid G=co) \exp(Z_i \alpha+X_i^\top \beta)\big\} \\ & \quad \times \big\{1-\zeta_{1,\xi}\big( F_{T \mid Z,X,G,\mu,\Lambda}(Y_i\mid Z_i,X_i,co),F_{C \mid Z,X,G,\eta}(Y_i\mid Z_i,X_i,co)\big)\big\}\Big]^{\Delta_{1,i}} \\ & \quad \times \Big[f_{C \mid Z,X,G,\eta}(Y_i \mid Z_i,X_i,co) \\ & \quad \times \big\{1-\zeta_{2,\xi}\big(F_{T \mid Z,X,G,\mu, \Lambda}(Y_i \mid Z_i,X_i,co) , F_{C \mid Z,X,G,\eta}(Y_i \mid Z_i,X_i, co)\big)\big\}\Big]^{\Delta_{2,i}}.
\end{align*}
Note that the distribution of $A$ can be omitted from the likelihood function due to Assumption \ref{Anoninfcens}. However, maximizing this likelihood directly is impossible as $G$ is a latent variable. Moreover, even if we were to observe $G$, maximizing this likelihood still poses a challenge as $\Lambda$ is an unknown function. To solve the problem of not observing $G$, we can use Theorem 3.1 by \citet{ABADIE2003231} as described in Section \ref{identcomp}. However, it is important to note that $\omega$ being negative when $Z \neq W$ can cause non-convexity issues to arise during the maximization of the weighted log-likelihood. Moreover, when we define our estimator for $\Lambda$, it will be clear that the negativity of $\omega$ can cause our estimated baseline cumulative hazard for the compliers to be non-monotonic. Therefore, we follow the approach by \citet{abadie2002instrumental} and replace $\omega$ by $\EX[\omega \mid S]$, where $\EX[\omega \mid S]$ is given by
  $$
  \kappa^*(S) = \EX[\omega \mid S]=\mathbb{P}(G = co \mid S=s) = 1-\frac{Z\mathbb{P}(W=0\mid S)}{\mathbb{P}\big(W = 0 \mid X\big)}-\frac{(1-Z)\mathbb{P}(W = 1\mid S) }{\mathbb{P}\big(W = 1 \mid X\big)},
  $$
  with $S=(Y,\Delta_1,\Delta_2,Z,X^\top)^\top$. It is shown in Section B of the Supplementary Material that $\kappa^*(S)$ is indeed the probability of being a complier conditional on $S$, which is therefore bounded between $0$ and $1$. In what follows, we start by proposing an estimator for the conditional probability of an individual being a complier given $S$. Secondly, we will replace the unknown function $\Lambda$ with an estimator $\hat{\Lambda}_{\theta}$ for given values of the finite-dimensional parameter $\theta$. This way, we can estimate $\theta$  by solving the score equations that are derived from the pseudo-profile likelihood function $L(\theta,\hat{\Lambda}_{\theta})$. 

\subsection{Estimating the probability of being a complier}\label{probabilityofcomplier}

We propose a non-parametric estimator of $\kappa^*(\cdot)$ as described by \citet{wei2021estimation}. Firstly, let $R=(Y,X^\top),$ $ \pi(X)=\mathbb{P}(W=1\mid X),$ $ \nu(S)= \mathbb{P}(W=1\mid S)$ and 
$$
\nu_{j,l,k}(R)= \mathbb{P}(W=1\mid R,\Delta_1 = j,\Delta_2 = l,Z=k), \quad \text{with} \quad j,l,k=0,1.
$$
From this, it follows that
$$
\nu(S_i)=\sum^1_{j=0} \sum^1_{l=0} \sum^1_{k=0}\mathbbm{1}(\Delta_{1,i} = j,\Delta_{2,i}=l, Z_i=k)\nu_{j,l,k}(R_i).
$$
Further, let $K_{h_{1,n}}(x) = h_{1,n}^{-m}K_1(\frac{x}{h_{1,n}})$ and $K_{h_{2,n}}(r)= h_{2,n}^{-(m+1)}K_2(\frac{r}{h_{2,n}})$ with $K_1$ and $K_2$ (multiplicative) kernel functions. Note that we have assumed that all the covariates in $X$ are continuous (otherwise we stratify on the discrete ones) such that we can estimate $\pi(x)$ and $\nu_{j,l,k}(r)$ by
$$
\hat{\pi}(x) = \frac{\sum_{i=1}^n K_{h_{1,n}}(x-X_i)W_i}{\sum_{i=1}^n K_{h_{1,n}}(x-X_i)}$$ and $$\hat{\nu}_{j,l,k}(r) = \frac{\sum_{i=1}^n \mathbbm{1}(\Delta_{1,i} = j,\Delta_{2,i}=l, Z_i=k)K_{h_{2,n}}(r-R_i)W_i}{\sum_{i=1}^n \mathbbm{1}(\Delta_{1,i} = j,\Delta_{2,i}=l, Z_i=k)K_{h_{2,n}}(r-R_i)}.
$$
From this, we estimate $\nu(S_i)$ by $\hat{\nu}(S_i)=\sum^1_{j=0} \sum^1_{l=0} \sum^1_{k=0}\mathbbm{1}(\Delta_{1,i} = j,\Delta_{2,i}=l, Z_i=k)\hat{\nu}_{j,l,k}(R_i)$. A non-parametric estimator of $\kappa(S_i)$ is then given by
$$
\hat{\kappa}(S_i)= 1-\frac{Z_i\big(1-\hat{\nu}(S_i)\big)}{1-\hat{\pi}(X_i)}-\frac{(1-Z_i)\hat{\nu}(S_i) }{\hat{\pi}(X_i)}.
$$
Note that $\hat{\kappa}(S_i)$ is a probability and should therefore be bounded between 0 and 1. Therefore, it is proposed to replace $\hat{\kappa}(S_i)$ by $\Tilde{\kappa}(S_i) = \min(\max(\hat{\kappa}(S_i), a_{l,n}),a_{u,n})$, where $a_{l,n}$ and $a_{u,n}$ are positive sequences that, as $n$ increases, approach 0 and 1 respectively.

\subsection{Estimating the baseline cumulative hazard function}\label{estimatingchf}

The non-parametric estimator for $\Lambda(\cdot\mid G=co)$ will be constructed using martingale ideas from Theorem 1.3.1 by \citet{fleming2011counting}. Firstly, let $I_i(y)=\mathbbm{1}(Y_i \leq y, \Delta_{1,i}=1)$ and $\tilde{I}_i(y)=\mathbbm{1}(Y_i \geq y)$. Following a similar martingale construction as \citet{Rivest2001AMA}, we have that
$$
\mathbbm{M}_i(y)=I_i(y) - \int^y_0 \tilde{I}_i(u)  \lambda^{\#} (u\mid Z_i,X_i,co)\diff u,
$$
are right-continuous martingales with respect to the $ \sigma$-algebra $$\mathcal{F}^{co,i}_y=\sigma\left\{I_i(u), \tilde{I}_i(u), Z_i,X_i, G = co:  0 < u < y \leq \bar{\tau} \right\},$$ where $\bar{\tau}$ is the maximum follow-up time defined in Assumption \ref{positiveprobassumptionwithtau} and $\lambda^{\#} (u\mid z,x,co)$ the conditional crude hazard rate, that is,
$$
\lambda^{\#} (y\mid z,x,co) = \frac{-\frac{\partial}{\partial u}\mathbbm{P}(T \geq u, C \geq y \mid Z = z, X = x, G = co)\mid_{u=y}}{\mathbbm{P}(T \geq y, C \geq y \mid Z = z, X = x, G = co)},
$$
for all $(z,x^{\top})$. Using the model specification from Section \ref{modelspec}, it can easily be shown that \begin{align*}
\lambda^{\#} (y\mid Z_i,X_i,co) = \exp\big(\psi_i(\theta^*,\Lambda^*(y\mid G=co))\big)\lambda^*(y\mid G = co),
\end{align*}
with
\begin{align*}
    & \psi_i(\theta^*,\Lambda^*(y\mid G=co)) \\ & =   Z_i\alpha^*+X_i^\top \beta^*-\Lambda^*(y\mid G=co)\exp(Z_i\alpha^*+X_i^\top \beta^*) -\log\left\{S_{\theta^*,\Lambda^*}(y \mid Z_i,X_i,co)\right\} \\ & \quad +\log\left\{1-\zeta_{1,\xi^*}\big(F_{T \mid Z,X,G,\mu^*,\Lambda^*}(y \mid Z_i,X_i,co),F_{C \mid Z,X,G,\eta^*}(y \mid Z_i,X_i,co) \big)\right\},
\end{align*}
and where $\theta^* = (\mu^*,\eta^*,\xi^*), $ $\Lambda^*$ and $\lambda^*$ are the true values of $\theta,\Lambda$ and $\lambda$ respectively. If we let $I_i(y-)=\lim_{u \text{ } \uparrow \text{ } y}I_i(u)$, it follows that $\diff I_i(y) = I_i(y)-I_i(y-)$ is a binary random variable with conditional probability $\tilde{I}_i(y)\exp\left(\psi_i(\theta^*,\Lambda^*(y \mid G=co))\right)\lambda^*(y \mid G=co)$ of being one given $\mathcal{F}^{co,i}_{y}$. Motivated by $\mathbbm{M}_i(y)$ being a right-continuous martingale, it follows that 
\begin{equation*}
\EX \Big[\diff I_i(y)-\tilde{I}_i(y)\exp\left(\psi_i\left(\theta^*,\Lambda^*(y \mid G=co)\right)\right)\diff \Lambda^*(y\mid G=co) \mid \mathcal{F}^{co,i}_{y} \Big] = 0, \end{equation*} with $0 < y \leq \bar{\tau}$ and $\Lambda^*(0\mid G=co)=0.$ However, this equation cannot be used for estimation since it depends on a latent variable $G$. Using Theorem 3.1 from \citet{ABADIE2003231} that we described before, we can rewrite this as
$$
\EX \bigg[ \kappa^*(S_i) \times \Big\{ \diff I_i(y)-\tilde{I}_i(y)\exp\left(\psi_i\left(\theta^*,\Lambda^*(y \mid G=co)\right)\right)\diff \Lambda^*(y \mid G = co) \Big\} \mid \mathcal{F}^i_{y} \bigg] = 0,
$$
with $\kappa^*$ the true value of $\kappa$ and $\mathcal{F}^i_{y}=\sigma\left\{I_i(u), \tilde{I}_i(u), Z_i,X_i: 0 < u < y \leq \bar{\tau}\right\}.$ In this way, the expectation no longer depends on the unobserved variable $G$. Therefore, for a given $\theta$, we could estimate $\Lambda^*(y \mid G=co)$  by solving the following functional estimating equation:
\begin{equation*}
n^{-1}\sum^n_{i=1} \bigg[ \tilde{\kappa}(S_i) \times \Big\{ \diff I_{i}(y)-\tilde{I}_{i}(y)\exp\left(\psi_i(\theta,\Lambda(y \mid G = co))\right)\diff \Lambda(y \mid G = co) \Big\} \bigg] = 0,    
\end{equation*}
with $0 < y \leq \Bar{\tau}$ and $\Lambda(0\mid G=co)=0.$
It is clear that $\diff I_{i}(y)$ will be equal to 1 at the observed event times $(Y_i, \Delta_{1,i}=1)$ and 0 everywhere else. Combining this with the fact that $\Tilde{\kappa}(S_i)$ is bounded between $0$ and $1$, it is clear that the estimator of $\Lambda^*$ following from this estimating equation is a non-decreasing step function where the jumps are at the ordered observed survival times: $0 = t_0 < t_1 < t_2 < \dots < t_K \leq \bar{\tau}$. However, using this estimating equation would involve a complex iterative optimization process that solves a $K$-dimensional system of equations, that is
$$
\hat{\Lambda}_{\theta}(t_k \mid G = co)=\hat{\Lambda}_{\theta}(t_{k-1} \mid G = co)+\frac{\sum^n_{i=1} \left[ \tilde{\kappa}(S_i) \times \diff I_{i}(t_k) \right]}{\sum^n_{i=1} \left[\tilde{\kappa}(S_i) \times \tilde{I}_{i}(t_k)\exp(\psi_i(\theta,\hat{\Lambda}_{\theta}(t_k\mid G = co)))\right]},
$$
for $k=1,\dots,K$ and with $\hat{\Lambda}_{\theta}(t_0 \mid G = co)=0$. Following \citet{ZuckerDavidM2005APLM}, we propose estimating $\Lambda^*$ as a step function with jumps at the ordered observed event times in the following way:
$$
\Delta \hat{\Lambda}_{\theta}(t_k \mid G = co)=\frac{\sum^n_{i=1} \left[ \tilde{\kappa}(S_i) \times \diff I_{i}(t_k) \right]}{\sum^n_{i=1} \left[\tilde{\kappa}(S_i) \times \tilde{I}_{i}(t_k)\exp(\psi_i(\theta,\hat{\Lambda}_{\theta}(t_{k-1}\mid G = co)))\right]},
$$
with $\lim_{y \text{ } \uparrow \text{ } t_1}\hat{\Lambda}_{\theta}(y \mid G = co)=0$ and $\Delta \hat{\Lambda}_{\theta}(t_k \mid G = co) = \hat{\Lambda}_{\theta}(t_k \mid G = co) - \hat{\Lambda}_{\theta}(t_{k-1} \mid G= co)$. Therefore, the estimates $\hat{\Lambda}_{\theta}(t_1 \mid G = co),\dots,\hat{\Lambda}_{\theta}(t_K \mid G = co)$ can be obtained by a simple forward recursion. When the independence copula is specified, it is important to note that $\hat{\Lambda}_{\theta}$ cancels out from the denominator. This results in the estimator simplifying to a weighted version of the \citet{breslow1974covariance} estimator. Moreover, the presence of $\hat{\Lambda}_{\theta}$ in the denominator when the specified copula deviates from the independence copula and the presence of $\tilde{\kappa}$ poses significant challenges in deriving the asymptotic properties of the estimator.

\subsection{Estimating the parameters of interest}\label{sectionzest}

We set the expected value (conditional on $G=co$) of the first-order conditions (with respect to $\theta$) of the conditional likelihood evaluated at the true $\theta$ and $\Lambda$ to zero to arrive at the following equations:
\begin{equation*}
\EX \left[ U\left(S_i,\theta^*,\Lambda^*\right) \mid G = co\right] = 0, 
\end{equation*}
where 
$$
U\big(S_i ,\theta^*,\Lambda^*\big) = \frac{\partial}{\partial \theta} \log f_{Y,\Delta_1,\Delta_{2} \mid Z,X,G,\theta^*,\Lambda^*}(Y_i,\Delta_{1,i},\Delta_{2,i} \mid Z_i,X_i,co),
$$
Again, we can use Theorem 3.1 from \citet{ABADIE2003231} to rewrite equation this in the following way:
\begin{equation*}
\EX \Big[ \kappa^*(S_i) \times  U\big(S_i,\theta^*,\Lambda^*\big)\Big] = 0,  
\end{equation*}
such that the score functions no longer depend on the unobserved variable $G$. Therefore, let $$M_n\big(\kappa,\theta,\Lambda\big) = n^{-1}\sum_{i=1}^n m\big(S_i,\kappa,\theta,\Lambda\big),$$ and $$m : \mathcal{S} \times\mathcal{K}\times \Theta \times \mathcal{L} \to \mathbb{R}^{\text{dim}(\Theta)} : \big(S,\kappa,\theta,\Lambda\big) \mapsto \kappa(S) \times U\big(S,\theta,\Lambda\big),$$ a measurable vector-valued function. Note that $\Theta$ is the parameter space for $\theta$ and $\mathcal{K}, \mathcal{L}$ are function spaces that will be defined in Section \ref{asymptotics}. Further, let $M(\kappa, \theta,\Lambda) = \EX[m(S,\kappa, \theta,\Lambda)] $. Using what we have derived so far, we have the following estimating equations:
\begin{equation*}
M_n(\tilde{\kappa},\theta,\hat{\Lambda}_{\theta}) = 0,    
\end{equation*} 
where $\hat{\theta}$ is defined as the solution to these weighted score equations. Note that this $Z$-estimator is only used for the theory, as in practice we will maximize the weighted logarithm of $L(\theta,\Lambda)$ with $\tilde{\kappa}(S_i)$ as the weights for each $i$. We do this by randomly generating $J$ starting values for $\theta$, denoted by $\tilde{\theta}_j$ with $j=1,\dots,J$, from a specified parameter space and calculating $\hat{\Lambda}_{\tilde{\theta}_j}$ for each $j$. The weighted log-likelihood is then maximized with respect to $\theta$ for each $j$, given $\hat{\Lambda}_{\tilde{\theta}_j}$, such that we have a set of estimates $\{\hat{\theta}_j\}_{j=1,\dots,J}$. Next, we continue with the $\hat{\theta}$ that has the highest weighted log-likelihood, and iterate calculating $\hat{\Lambda}_{\hat{\theta}}$ and maximizing the weighted log-likelihood until a convergence criterion is met or a maximum amount of iterations is reached.

\section{Asymptotic properties}\label{asymptotics}

In this section, it will be shown that the finite-dimensional parameter estimates $\hat{\theta}$ resulting from the $Z$-estimator defined in Section \ref{sectionzest} are consistent and asymptotically normal. The difficulty in establishing the asymptotic theory comes from our estimating equations involving both finite and infinite-dimensional parameters. This is complicated further by the fact that $\hat{\Lambda}_{\theta}$  depends on $\tilde{\kappa}$. Firstly, let 
\begin{align*}
\Gamma_{\theta} = \Gamma_{\theta}(\kappa^*,\theta^*,\Lambda_{\theta^*}^*) & = \EX\left[\frac{\partial}{\partial \theta^\top} m(\kappa^*, \theta,\Lambda_{\theta^*}^*)\bigg\rvert_{\theta=\theta^*}\right] \\ & \quad + \EX\left[\frac{\partial}{\partial \Lambda} m(\kappa^*, \theta^*,\Lambda)\bigg\rvert_{\Lambda = \Lambda^*_{\theta^*}}\times\frac{\partial}{\partial \theta^\top} \Lambda^*_{\theta}(Y \mid G = co)\bigg\rvert_{\theta=\theta^*}\right].
\end{align*}
In addition to the previously made assumptions, we need the following regularity conditions to establish the asymptotic properties of the estimator:

\begin{enumerate}[label=(C\arabic*),resume,left=0.25\leftmargin]

\item Let $X_c$ be the subvector of $X$ containing the continuous covariates and $m_c$ the dimension of $X_c$. The discrete components of $X$ take on finite values and $(Y,X^\top_c)$, conditional on $(Z,\Delta_1,\Delta_2)$ and the discrete components of $X$, has a support equal to a product of compact intervals with a density that is continuously differentiable of the order $b$ and bounded away from zero and infinity. \label{boundedsupportassumption}

\item For some $a_0 > 0,$ $ \kappa^*(S) > a_0$ almost surely and for some $0<a_1<a_2<1,$ $ a_1 < \pi^*(X) < a_2$ almost surely, where $\pi^*$ is the true value of $\pi$.\label{boundedprobassumption}

\item Let $K_1$ and $K_2$ be symmetric kernels that are zero outside of a bounded set. There exists a positive integer $d$, with $2d > m_c+1$, such that $K_1$ and $K_2$ are continuously differentiable of order $d$. Moreover, for $k=1,2$, we have that $\int K_k(u)du = 1$ and for all $j < b,$ with $b$ defined by Condition \ref{boundedsupportassumption}, we have that $ \int K_k(u)[\bigotimes_{l=1}^j u]du = 0$, where $\bigotimes_{l=1}^j u$ stands for executing $j$ times the Kronecker product on $u$.\label{kernelassumption} 

\item $\nu^*(\cdot)$ and $\pi^*(\cdot)$ are at least $p$-th order continuously differentiable with $p$ an integer such that $p \geq d+b$. Moreover, $$\frac{nh_{1,n}^{2m_c+4d}}{( \log n)^2 } \to \infty, \frac{nh_{2,n}^{2m_c+2+4d}}{( \log n)^2 } \to \infty \text{ and } nh_{k,n}^{2b} \to 0 \text{ as } n \to \infty.$$
 \label{bandwithassumption}

\item $a_{l,n} = o(n^{-1/2})$ and $1-a_{u,n} = o(n^{-1/2})$.\label{boundedestimatassumption}

\item The parameter space $\Theta$ is compact and contains an open neighborhood of the true parameter vector $\theta^*$. \label{compactparspaceassumption}

\item The function $\Lambda^*(t \mid G=co)$ is monotone increasing and differentiable with derivative $\lambda^*(t \mid G=co)$. In addition, $\lambda^*(t \mid G=co)$ is bounded from above by some constant $\Bar{\lambda}$ for all $t \in [0,\bar{\tau}].$\label{assumptionsoncumhazard}

\item  The functions $S(t \mid z,x,co), \zeta_{1}\big(F_{T \mid Z,X,G}(t\mid z,x,co) , F_{C \mid Z,X,G}(t \mid z,x,co)\big)$ and \\$\zeta_{2}\big(F_{T \mid Z,X,G}(t\mid z,x,co) , F_{C \mid Z,X,G}(t \mid z,x,co)\big)$ exist and are twice continuously differentiable with respect to $\theta$ over $\Theta$. Also, all derivatives of order two (with respect to $\theta$) are bounded, uniformly in $\Lambda,\kappa,t,z,x$ over their relevant domain.  \label{continuousderivassumption}

\item $\Gamma_{\theta} $ is a finite and nonsingular matrix. \label{fullrankass}
\end{enumerate}
Condition \ref{boundedsupportassumption} implies that the support of $X$, denoted by $\mathcal{X} \subset \mathbb{R}^m$ is bounded. Moreover, the densities of $Y$ or $X$ are also bounded and positive. Note that Assumption \ref{positiveprobassumptionwithtau} and Conditions \ref{boundedsupportassumption} and \ref{continuousderivassumption} imply that $\exp\big(\psi(\theta,c)\big)$ is bounded on the interval $[\Psi_{\min},\Psi_{\max}]$ over $\theta \in \Theta, s \in \mathcal{S} $ and for all $ c \in \mathbb{R}_{\geq 0}$ and that Condition \ref{boundedprobassumption} implies that $\kappa^*(S)$ and $\pi^*(X)$ are bounded away from 0 and 1 almost surely. Moreover, Condition \ref{kernelassumption} implies that $h_{1,n} = o(n^{-1/2b} \wedge (\log n)^{\nicefrac{1}{(m_c+2d)}}n^{\nicefrac{-1}{(2m_c+4d)}})$ and $ h_{2,n} = o(n^{-1/2b} \wedge (\log n)^{\nicefrac{1}{(m_c+1+2d)}}n^{\nicefrac{-1}{(2m_c+2+4d)}})$. Further, we define $\mathcal{K}$ to be the class of all functions $\kappa(\cdot)$, defined over $\mathcal{S}$, which are bounded between 0 and 1 and that are at least $d$-th order continuously differentiable, where $d$ is defined by Condition \ref{kernelassumption}. Further, let $\mathcal{L}$ be the class of all functions $\Lambda(\cdot)$, defined over $[0,\bar{\tau}]$, for which $\Lambda(0)=0$, $\Lambda(\cdot)$ is non-decreasing, and $\Lambda(\bar{\tau}) < \Lambda_{\max}$, where $\Lambda_{\max}$ is defined by Lemma 3 in Section C of the Supplementary Material. We are now ready to state the following theorems, of which the proof can be found in Section D of the Supplementary Material.

\begin{theorem}\label{consisttheorem}
If Assumptions \ref{AindepIV}-\ref{positiveprobassumptionwithtau} hold true and Conditions \ref{identifiabilityC}-\ref{fullrankass} are satisfied, it follows that
$$ \hat{\theta}\xrightarrow{\text{ p }}\theta^*.$$
\end{theorem}

\begin{theorem}\label{assntheorem}
If Assumptions \ref{AindepIV}-\ref{positiveprobassumptionwithtau} hold true and Conditions \ref{identifiabilityC}-\ref{fullrankass} are satisfied, it follows that there exists a positive definite matrix $\Omega$ such that
$$
n^{1/2}(\hat{\theta}-\theta^*) \xrightarrow{\text{ d }} \mathcal{N}(0,\Sigma),
$$
where $\Sigma=\Gamma_{\theta}^{-1}\Omega\left(\Gamma_{\theta}^{-1}\right)^{\top}$.
\end{theorem}

To show the asymptotic properties of $\hat{\theta}$, we use results by \citet{chen2003estimation}. However, to prove Theorem \ref{assntheorem}, we need to extend their results to the case where the criterion function depends on two unknown functions, since $M_n$ depends on both $\tilde{\kappa}$ and $\hat{\Lambda}_{\theta}$. Checking the conditions of this extended version of the theorem is complicated further by $\hat{\Lambda}_{\theta}$ depending on $\Tilde{\kappa}$. Note that $\Omega$ has a very lengthy expression that, since we estimate $\Sigma$ through a naive bootstrap method, is of little use and therefore omitted.

\section{Simulation results}\label{sectsimresul}

In this section, various simulation studies are performed to investigate the finite sample performance of the proposed estimator. Firstly, we look at the performance of our estimator using multiple combinations of copulas and marginals. The proposed estimator is compared to two other estimators: one that does not account for unobserved heterogeneity (naive estimator) and one that uses the proposed method when $G$ is observed (oracle estimator). Secondly, we assess the performance of the proposed estimator under misspecification of the censoring distribution or the copula model. Lastly, we look at what happens to our estimates when we change either the proportion of compliers or the sample size.

\subsection{Comparison of the proposed estimator}
The first steps of the data-generating process, which are the same for all designs, are as follows:
$$
X = (X_1,X_2)^\top \text{ where } X_1 \sim \text{Bernoulli}(0.5) \text{ and } X_2 \sim U(0,1).
$$
Further, we generate $G$ from a multinomial distribution with $\mathbb{P}(G=co)=2/3$ and $\mathbb{P}(G=at) = \mathbb{P}(G=nt)=1/6.$ Moreover, $W$ is generated from a Bernoulli$(\pi(X))$ distribution, where
$$
\pi(X) = \frac{\exp\left(0.5X_1 + X_2 + 2X_1X_2 + \epsilon\right)}{1+\exp\left(0.5X_1 + X_2 + 2X_1X_2 + \epsilon\right)},
$$
with $\epsilon \sim \mathcal{N}(0,0.25^2).$ Note that by adding this $\epsilon$, a logistic regression model for $\pi(X)$ would be misspecified. It follows that we can determine $Z$ by using $G$ and $W$. The next step depends on the choice of the copula and censoring distribution. As an example, we consider the following design (Frank - Weibull):
$$
\mathbb{P}(T \leq t, C\leq c\mid Z=z,X=x,G=g)=\mathcal{C}_{\xi_g}\left(F_{T \mid Z,X,G}(t\mid z,x,g),F_{C \mid Z,X,G}(c\mid z,x,g)\right),$$
with $\mathcal{C}_{\xi_g}$ a Frank copula, that is, 
$$
\mathcal{C}_{\xi_g}(u,v) = -\frac{1}{\xi_g}\log\left(1+\frac{\left(\exp(-{\xi_g} u)-1\right)\left(\exp(-{\xi_g} v)-1\right) }{\exp(-{\xi_g})-1}\right), \quad {\xi_g} \neq 0.
$$
Furthermore, we specify the distribution of $T$ and $C$ by a proportional hazards and Weibull model respectively, that is,
$$
F_{T \mid Z,X,G}(t\mid z,x,g)=1-\exp\{-\Lambda(t\mid G=g) \exp( z \alpha_g+x^\top \beta_g)\},
$$
$$
F_{C \mid Z,X,G}(c\mid z,x,g) = 1 - \exp\left(-\exp\left(\frac{\log(c) - \tilde{x}^{\top}\eta_g}{\nu_g}\right)\right),
$$
where $\tilde{x} = (1,z,x^{\top})^{\top}$, $\beta_g^\top = (\beta_{1,g},\beta_{2,g})$ and $\eta_{g}^{\top} = (\eta_{0,g},\eta_{1,g},\eta_{2,g},\eta_{3,g},\nu_{g})$. Note that specifying the association parameter $(\xi_g)$ or Kendall's tau  $(\tau_g)$, is equivalent because of the following relation:
$$
\tau_g = 4\int_0^1\int_0^1\mathcal{C}_{\xi_g}(u,v)\textit{c}_{\xi_g}(u,v)dudv-1,
$$ which is one-to-one for the (rotated) Archimedean and Gaussian copulas. This approach maintains the same model structure across all latent subgroups while allowing the parameter values to vary. Examples of this can be seen in Tables \ref{designs1} and \ref{designs2}, which represent a low and high dependence scenario respectively.
\begin{table}[H]
    \centering
        \caption{Parameter specification for a scenario with low dependence.}
    \renewcommand{\arraystretch}{1.25}
    \begin{tabular}{|c|c|c|}
    \hline
     Low dependence & $\Lambda(t\mid G=g)$ & $(\tau_g,\alpha_g,\beta_{1,g},\beta_{2,g},\eta_{0,g},\eta_{1,g},\eta_{2,g},\eta_{3,g},\nu_{g})$ \\   \hline
     $g = co$ & $0.5t^{\nicefrac{3}{4}}$ & (0.25, -0.6, 1, 0.9, 1.5, -0.8, -2, 0.9, 1.2) \\   \hline
     $g = at,nt$  & $0.7t^{\nicefrac{3}{5}}$ & (0.2, -0.1, 0.8, 0.7, 1.3, -1, -1.8, 0.6, 1.1) \\   \hline
    \end{tabular}
    \renewcommand{\arraystretch}{1}
    \label{designs1}
\end{table}

\begin{table}[H]
    \centering
    \caption{Parameter specification for a scenario with high dependence.}
    \renewcommand{\arraystretch}{1.25}
    \begin{tabular}{|c|c|c|}
    \hline
     High dependence & $\Lambda(t\mid G=g)$ & $(\tau_g,\alpha_g,\beta_{1,g},\beta_{2,g},\eta_{0,g},\eta_{1,g},\eta_{2,g},\eta_{3,g},\nu_{g})$ \\   \hline
     $g = co$ & $0.1t^{\nicefrac{3}{5}}$ & (0.75, 0.6, 1.3, 1, 1.3, -0.6, -0.8, 1.2, 1) \\   \hline
     $g = at,nt$  & $0.2t^{\nicefrac{7}{10}}$ & (0.7, 0.2, 1.1, 0.8, 1.1, -0.8, -0.5, 0.9, 1.2) \\   \hline
    \end{tabular}
    \renewcommand{\arraystretch}{1}
    \label{designs2}
\end{table}
To simulate the administrative censoring, let $A \sim U(0,15)$ for the low dependence scenario and let $A \sim U(0,50)$ for the high dependence scenario. Finally, we can generate the follow-up time $Y$ = min\{$T,C,A$\}, $\Delta_1 = \mathbbm{1}(Y=T)$ and $\Delta_2 = \mathbbm{1}(Y=C)$.

The data-generating process was repeated 500 times with a sample size of 1000 for each of the six simulation designs considered. For each of the designs, we compare our proposed estimator to two others. The naive estimator ignores the endogeneity issue such that everyone is considered to be a complier ($\tilde{\kappa}(S_i) = 1$ for each $i$). The oracle estimator uses the same estimation procedure as the proposed estimator but treats $G$ as if it were observed ($\tilde{\kappa}(S_i) = \mathbbm{1}(G_i=co)$ for each $i$). Therefore, our proposed method reduces to a one-step estimation procedure as we do not need to estimate $\kappa^*$. We report the bias, empirical standard deviation (ESD), root mean squared error (RMSE) and coverage rate (CR). Note that the CR indicates the percentage of simulations in which the true parameter value falls within the estimated 95\% confidence interval. These coverage rates were computed using the warp-speed method described by \citet{warpspeed}. This method calculates bootstrap confidence intervals by drawing only one bootstrap resample per Monte Carlo sample, significantly reducing computation time. However, this approach may result in less accurate coverage rates due to the reduced number of bootstrap resamples. To estimate the probability of being a complier, we modified the {\fontfamily{cmtt}\selectfont point.est.kernel} function from the {\fontfamily{cmtt}\selectfont CCQTEC} package \citep{wei2021estimation} in {\fontfamily{cmtt}\selectfont R} to fit our setting. We use a multiplicative sixth-order Epanechnikov kernel and select the bandwidths from the range $\{0.01, 0.02,\cdots, 1\}$ using 10-fold cross-validation, with $a_{l,n} = 10n^{-1}$ and $a_{u,n} = 1-10n^{-1}$. In the second step, to estimate the parameters of interest, we maximize the weighted log-likelihood as explained in Section \ref{sectionzest} with $J=100$ and a maximum amount of iterations equal to 120.

The results for the scenario with low dependence can be found in Table \ref{tableestresults}. Each of these three designs has around 30 to 40 percent dependent and 5 to 10 percent administrative censoring. Note that the design in the middle column is based on a data-generating process with negative dependence. The results show that our method has low bias across all designs, especially compared to the naive estimator where the bias is the highest for the main parameter of interest $\alpha$. Compared to the naive estimator, the proposed estimator exhibits a substantially lower RMSE for the parameter of interest $\alpha$. However, its RMSE is higher for the other parameters. Nonetheless, the coverage rates of the naive estimator frequently show substantial deviations from the nominal 95\% level. In contrast, the coverage rates of the proposed estimator exhibit only minor deviations from the nominal level, likely due to the use of the warp-speed method. The results for the scenario with high dependence are similar and can be found in Section E of the Supplementary Material. 
\begin{table}[ht]
    \centering
    \caption{Estimation results for various designs with low dependence. Each design has around 30-40\% dependent and 5-10\% administrative censoring. Given are the bias, the empirical standard deviation (ESD), the root mean squared error (RMSE) and the confidence rate (CR).}
    \scalebox{0.8}{
\begin{tabular}{|c|crccc|rccc|rccc|}
 \hline
   \multirow{32}{*}{\rotatebox[origin=c]{90}{Low dependence scenario ($\tau = 0.25$)}} & \multicolumn{5}{c}{Frank - Weibull} & \multicolumn{4}{|c|}{Clayton(90) - log-normal} & \multicolumn{4}{c|}{Clayton(180) - log-logistic} \\
 \cline{2-14}
  & \multicolumn{13}{c|}{naive estimator} \\
 \cline{2-14}
  & & Bias & ESD & RMSE & CR & Bias & ESD & RMSE & CR & Bias & ESD & RMSE & CR \\ 
\cline{2-14}
& $\alpha$ & 0.294 & 0.105 & 0.312 & 0.342        & 0.299 & 0.106 & 0.317 & 0.140         & 0.303 & 0.105 & 0.321 & 0.226 \\ 
& $\beta_{1}$ & -0.125 & 0.098 & 0.159 & 0.972      & -0.111 & 0.119 & 0.163 & 0.802         & -0.129 & 0.087 & 0.155 & 0.660 \\
& $\beta_{2}$ & -0.184 & 0.167 & 0.248 & 0.874        & -0.155 & 0.155 & 0.220 & 0.802         & -0.193 & 0.151 & 0.244 & 0.876 \\ 
& $\eta_0$ & -0.038 & 0.216 & 0.219 & 0.996           & -0.097 & 0.240 & 0.259 & 0.934         & -0.157 & 0.349 & 0.383 & 0.978 \\ 
&  $\eta_1$ & -0.154 & 0.174 & 0.232 & 0.942        & -0.081 & 0.182 & 0.199 & 0.888         & -0.114 & 0.288 & 0.309 & 0.948 \\ 
& $\eta_2$ & 0.083 & 0.091 & 0.123 & 0.850          & 0.103 & 0.111 & 0.151 & 0.786         & 0.129 & 0.174 & 0.217 & 0.922 \\ 
&  $\eta_3$ & -0.047 & 0.187 & 0.193 & 0.966         & -0.096 & 0.225 & 0.244 & 0.898         & -0.118 & 0.339 & 0.359 & 0.966 \\ 
&  $\nu$ & -0.073 & 0.046 & 0.086 & 1.000           & -0.019 & 0.079 & 0.082 & 0.948         & -0.067 & 0.076 & 0.101 & 0.926\\ 
&  $\tau$ & 0.022 & 0.115 & 0.116 & 0.998         & -0.068 & 0.119 & 0.137 & 0.826         & 0.033 & 0.081 & 0.088 & 0.988\\ 
 \cline{2-14}
 & \multicolumn{13}{c|}{proposed estimator} \\
  \cline{2-14}
& $\alpha$ & 0.019 & 0.181 & 0.181 & 0.936        & 0.018 & 0.143 & 0.144 & 0.924         & -0.014 & 0.149 & 0.150 & 0.872 \\ 
& $\beta_{1}$ & -0.011 & 0.240 & 0.240 & 0.972      & 0.019 & 0.240 & 0.241 & 0.898         & -0.001 & 0.157 & 0.157 & 0.878 \\
& $\beta_{2}$ & -0.038 & 0.317 & 0.319 & 0.958        & -0.015 & 0.301 & 0.301 & 0.908         & -0.009 & 0.276 & 0.276 & 0.940 \\ 
& $\eta_0$ & 0.053 & 0.418 & 0.421 & 0.978           & 0.019 & 0.332 & 0.332 & 0.934         & 0.001 & 0.438 & 0.438 & 0.984 \\ 
&  $\eta_1$ & -0.035 & 0.288 & 0.290 & 0.980        & 0.016 & 0.222 & 0.223 & 0.924         & 0.048 & 0.351 & 0.353 & 0.966 \\ 
& $\eta_2$ & -0.040 & 0.175 & 0.180 & 0.888          & -0.025 & 0.236 & 0.238 & 0.922         & -0.014 & 0.345 & 0.345 & 0.952 \\ 
&  $\eta_3$ & 0.014 & 0.371 & 0.370 & 0.986         & -0.096 & 0.472 & 0.481 & 0.914         & -0.013 & 0.717 & 0.716 & 0.926 \\ 
&  $\nu$ & -0.058 & 0.082 & 0.100 & 0.966           & 0.008 & 0.106 & 0.107 & 0.962         & 0.006 & 0.094 & 0.095 & 0.982\\ 
&  $\tau$ & -0.006 & 0.273 & 0.273 & 0.996         & -0.029 & 0.149 & 0.152 & 0.994         & 0.022 & 0.111 & 0.113 & 0.998\\ 
   \cline{2-14}
 & \multicolumn{13}{c|}{oracle estimator} \\
 \cline{2-14}
 & $\alpha$ & 0.007 & 0.136 & 0.136 & 0.984        & 0.000 & 0.123 & 0.123 & 0.922         & 0.007 & 0.135 & 0.135 & 0.930 \\ 
& $\beta_{1}$ & 0.002 & 0.138 & 0.137 & 0.986      & 0.006 & 0.153 & 0.153 & 0.934         & -0.001 & 0.111 & 0.111 & 0.960 \\
& $\beta_{2}$ & -0.015 & 0.216 & 0.217 & 0.972        & 0.004 & 0.189 & 0.188 & 0.960         & -0.016 & 0.199 & 0.199 & 0.976 \\ 
& $\eta_0$ & -0.001 & 0.280 & 0.280 & 0.990           & 0.019 & 0.275 & 0.276 & 0.938         & -0.041 & 0.386 & 0.388 & 0.980 \\ 
&  $\eta_1$ & -0.007 & 0.217 & 0.217 & 0.992        & -0.005 & 0.206 & 0.206 & 0.966         & 0.007 & 0.323 & 0.322 & 0.954 \\ 
& $\eta_2$ & 0.002 & 0.120 & 0.119 & 0.976          & 0.004 & 0.147 & 0.146 & 0.910         & 0.014 & 0.234 & 0.235 & 0.956 \\ 
&  $\eta_3$ & 0.012 & 0.244 & 0.244 & 0.996         & -0.007 & 0.271 & 0.271 & 0.948         & 0.014 & 0.415 & 0.415 & 0.964 \\ 
&  $\nu$ & -0.007 & 0.063 & 0.063 & 0.980           & 0.009 & 0.097 & 0.097 & 0.932         & -0.012 & 0.088 & 0.088 & 0.966\\ 
&  $\tau$ & 0.005 & 0.171 & 0.171 & 1.000         & 0.004 & 0.127 & 0.127 & 0.986         & 0.014 & 0.093 & 0.094 & 0.992\\ 
\hline
\end{tabular}
}\label{tableestresults}
\end{table}

\subsection{Misspecification}

In this subsection, we evaluate the performance of the proposed method under misspecification of either the copula or the distribution of $C$. For each design, 500 datasets with a sample size of 1000 were generated. The parameter values used were those from the previously described low-dependence scenario. The Frank copula was used for all designs involving misspecification of the censoring distribution. When the copula was misspecified, a log-normal distribution was used for the censoring distribution. Detailed results on the impact of misspecification of the censoring distribution and copula are provided in Section E of the Supplementary Material. Overall, misspecification of the copula has a smaller effect on the bias and coverage rates compared to misspecification of the censoring distribution. However, this observation does not hold when the copula fails to model the correct direction of the dependence, as seen when data were generated using a Clayton(180) copula but a Clayton(90) copula was specified. The bias of the main parameter of interest, $\alpha$, remains relatively low across all designs, particularly compared to the naive estimator. For most designs, the bias and RMSE of $\alpha$ for the proposed estimator are comparable to those of the oracle estimator.

\subsection{Effect of the proportion of compliers and sample size}

\begin{figure}[H]
\caption{Effect of the complier ratio on the bias and RMSE of $\alpha$}
\label{compratioalpha}
\centering
\begin{subfigure}{.5\textwidth}
  \centering
  \includegraphics[width=\linewidth]{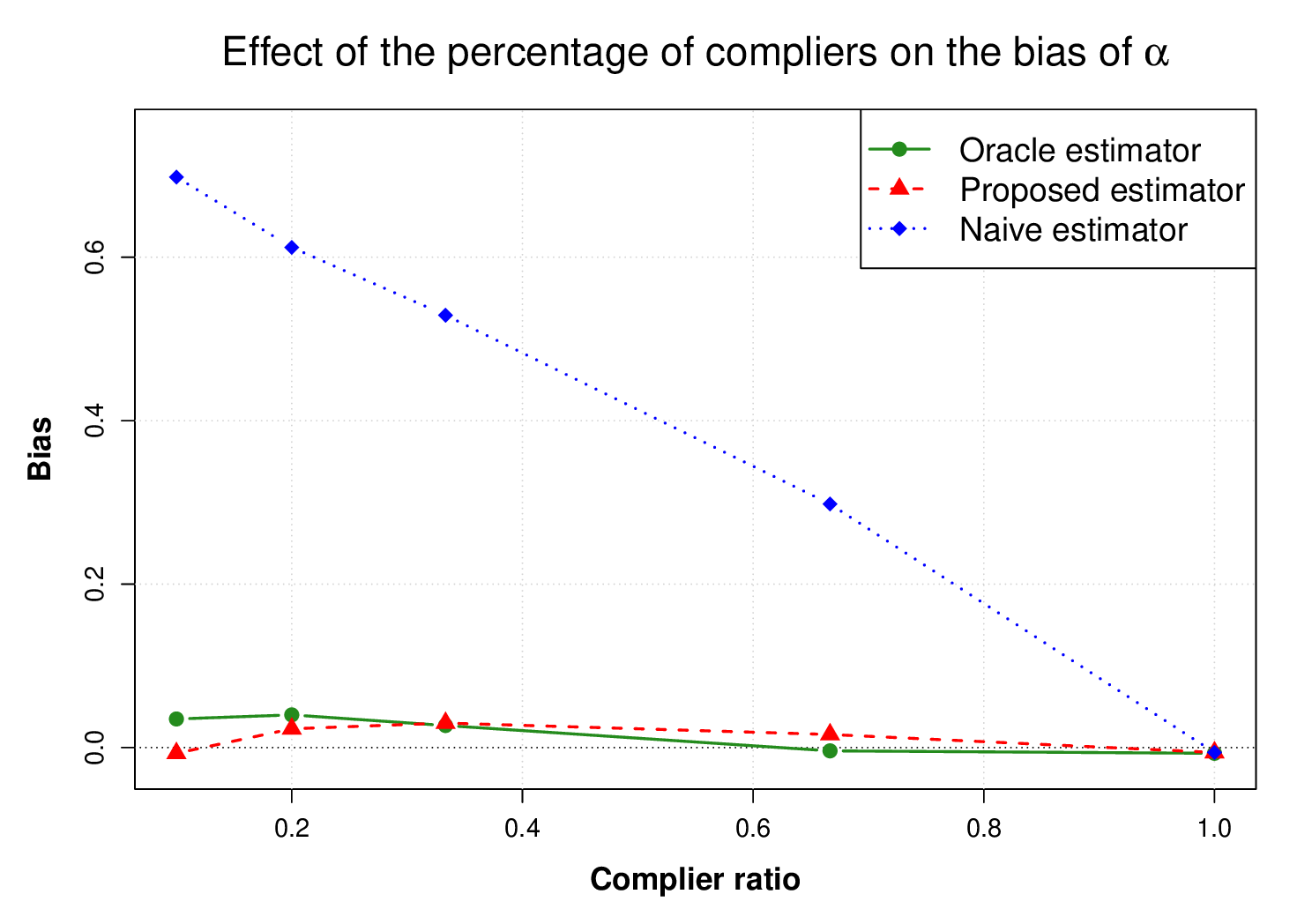}
\end{subfigure}%
\begin{subfigure}{.5\textwidth}
  \centering
  \includegraphics[width=\linewidth]{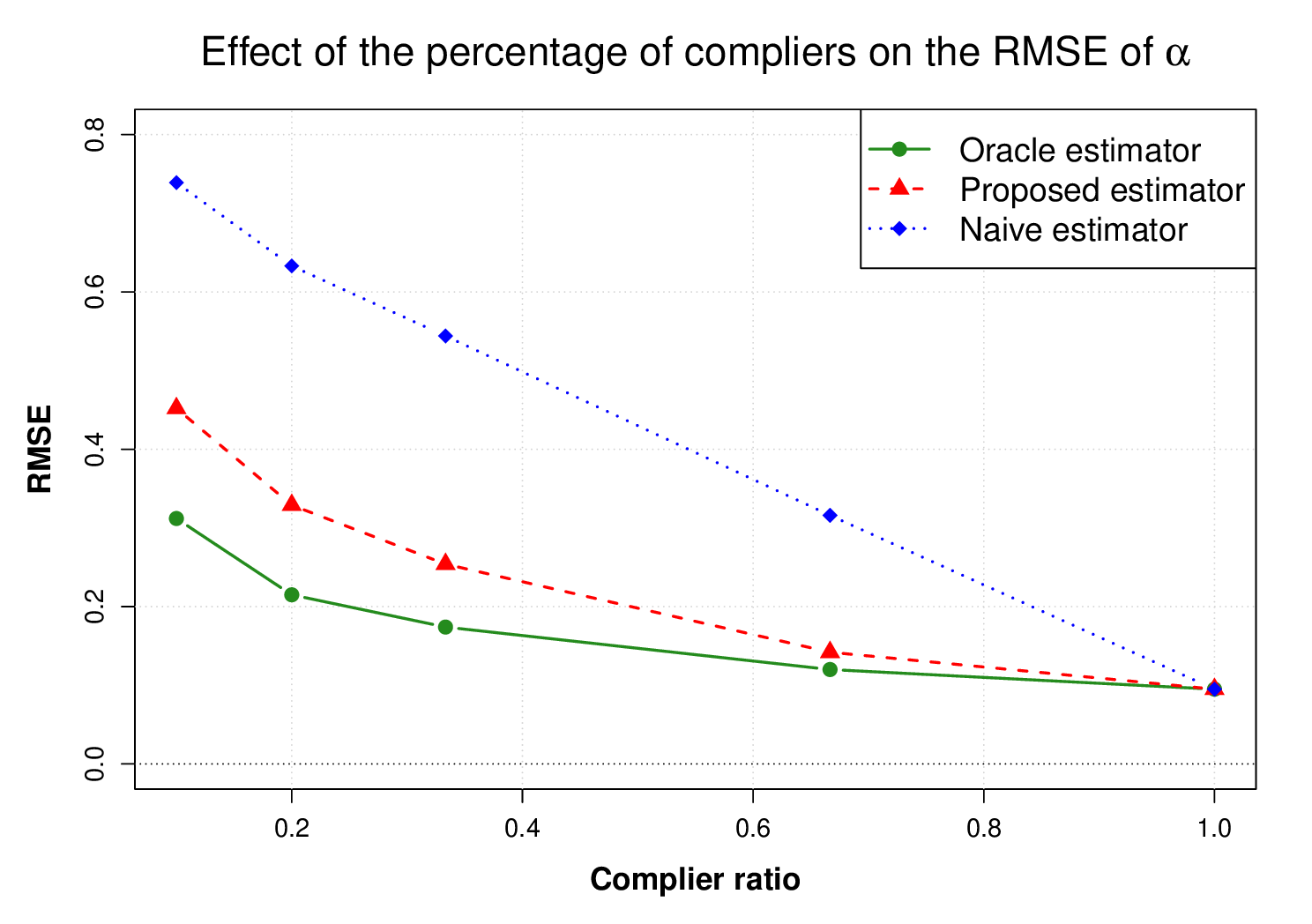}
\end{subfigure}
\end{figure}

\begin{figure}[H]
\caption{Effect of the sample size on the bias and RMSE of $\alpha$}
\label{sampsizealpha}
\centering
\begin{subfigure}{.5\textwidth}
  \centering
  \includegraphics[width=\linewidth]{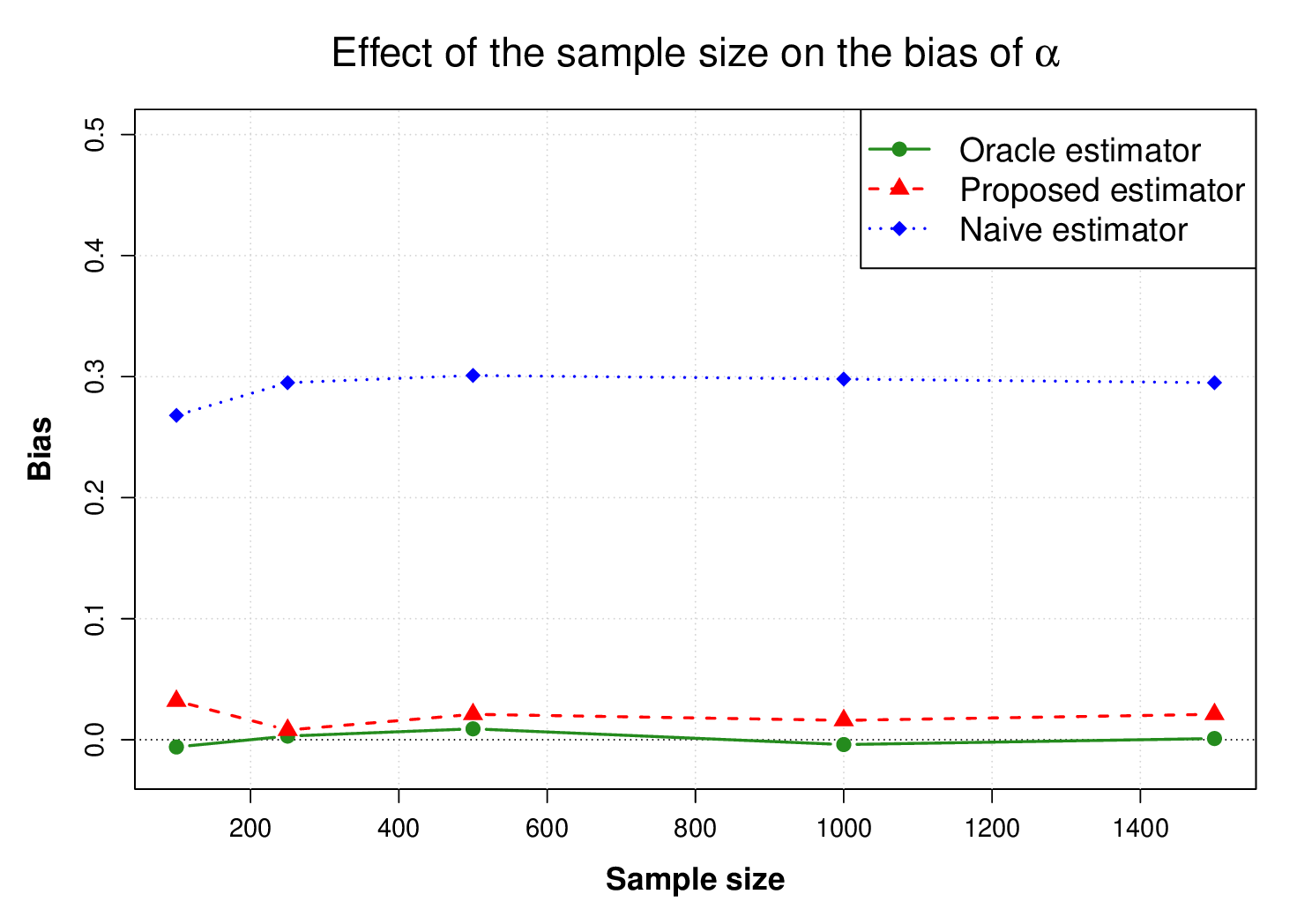}
\end{subfigure}%
\begin{subfigure}{.5\textwidth}
  \centering
  \includegraphics[width=\linewidth]{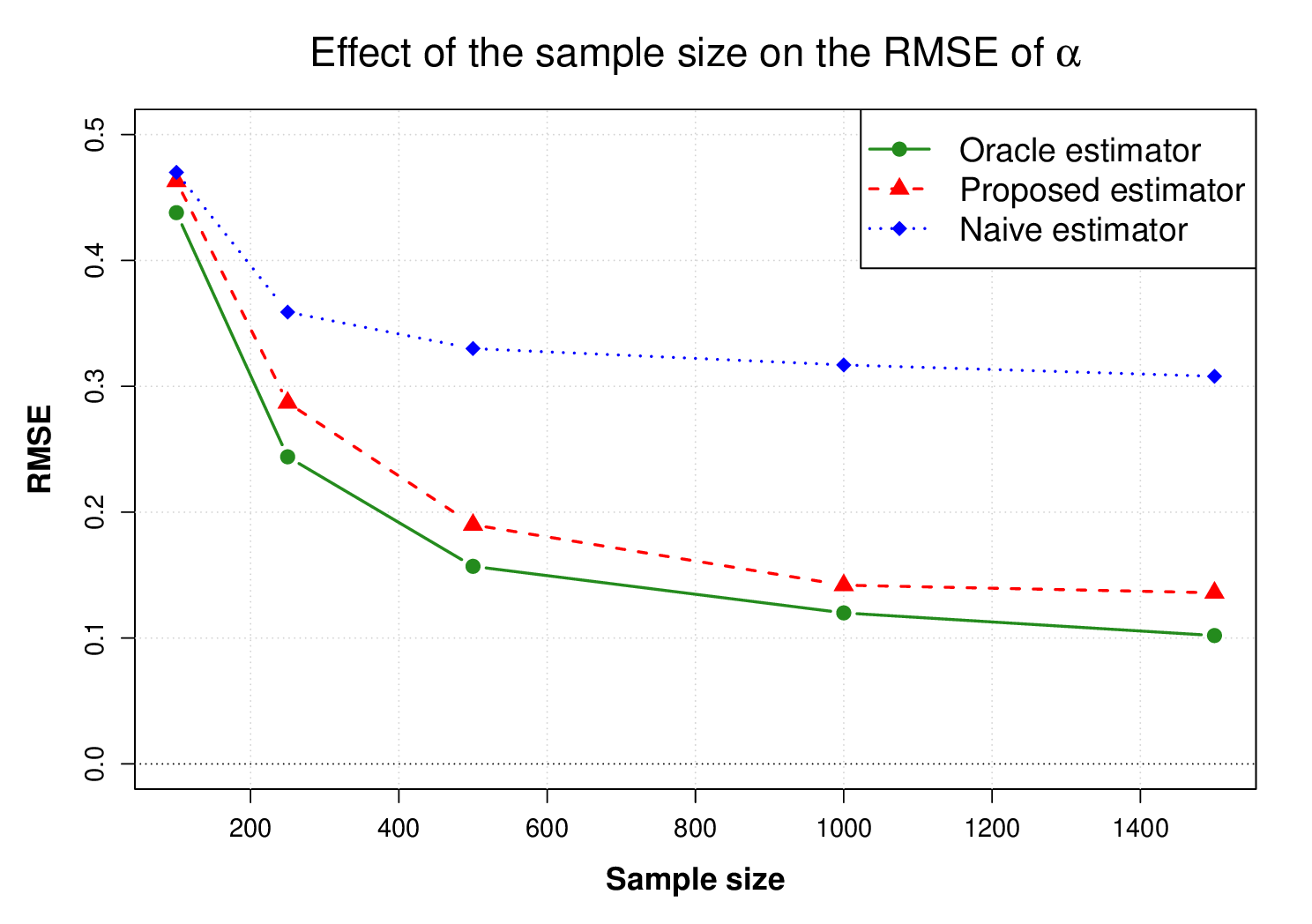}
\end{subfigure}
\end{figure}

Finally, we examine the impact of the proportion of compliers and sample size on the bias and RMSE of the parameter of interest $\alpha$. To investigate the effect of the proportion of compliers, we generate 250 datasets for 5 different complier ratios $(1/10, 1/5, 1/3, 2/3, 1)$ with a sample size of 1000 using a Clayton(90) copula and a log-normal censoring distribution. For the effect of the sample size we also generated 250 datasets for each of the 5 sample sizes $(100,250,500,1000,1500)$ with a complier ratio of 2/3 and using the same copula and censoring distribution. The parameter values used to generate the data sets were those from the previously described low-dependence scenario. The results for the effect of the complier ratio and the sample size can be found in Figures \ref{compratioalpha} and \ref{sampsizealpha} respectively. Our findings indicate that, regarding the bias, the proposed method is fairly robust to a low proportion of compliers. Even with only 10\% compliers, the proposed estimator exhibits a bias that is comparable to the oracle estimator. Moreover, the proposed estimator substantially outperforms the naive estimator in terms of both bias and RMSE regardless of the complier ratio. As the proportion of compliers increases to 100\% the bias and RMSE of all estimators converge, as would be expected. Regarding the effect of sample size, our proposed estimator maintains good performance with respect to bias, even when the sample size is as small as 100. In contrast, the bias of the naive estimator remains constant as the sample size increases. Furthermore, the RMSE of the naive estimator appears to plateau, showing little improvement with increasing sample size. Overall, these results suggest that our method performs effectively even in scenarios with a low proportion of compliers or small sample sizes, demonstrating its robustness under various conditions.

\section{Data Application}\label{sectdataap}

In this section, we present an empirical application of the proposed methodology to data from the Job Training Partnership Act (JTPA) study. A second data application uses data from the Health Insurance Plan of Greater New York (HIP) experiment and can be found in Section F of the Supplementary Material.

\subsection{The National Job Training Partnership Act Study}\label{JTPA}

The data examined come from a large-scale randomized experiment known as the National Job Training Partnership Act (JTPA) Study and have been analyzed extensively by \citet{bloom1997benefits}, \citet{abadie2002instrumental}, \citet{frandsen2015treatment}, \citet{wuthrich2020comparison}, \citet{beyhum2024instrumental} and \citet{crommen2024instrumental} among others. The data that we investigate is the same as in \citet{abadie2002instrumental} and \citet{wuthrich2020comparison}, but the problem of interest differs. This is because we investigate the effect of the job training services on unemployment duration, while they look at the effect of JTPA services on the sum of earnings in the 30-month period after treatment assignment. The problem investigated is more similar to \citet{frandsen2015treatment} and \citet{beyhum2024instrumental}, but again differs as we allow for dependent censoring. Lastly, \citet{crommen2024instrumental} analyzed the data under possible dependent censoring. However, their goal was to estimate a population treatment effect, while we focused on estimating a local treatment effect for the subgroup of compliers, which allows for more flexible modeling.

The JTPA study was conducted to assess the effectiveness of more than 600 federally funded programs established under the Job Training Partnership Act of 1982, aimed at improving the employability of eligible adults and out-of-school youths. These programs provided services such as classroom training, on-the-job training, and job search assistance. Funding for these initiatives began in October 1983 and continued into the late 1990s. Between 1987 and 1989, over 20000 adults and out-of-school youths who applied for JTPA were randomly assigned to either a treatment group, which was eligible for JTPA services, or a control group, which was not eligible for 18 months. However, due to some local program staff not strictly adhering to the randomization guidelines, about 3\% of the control group members received JTPA services. It is important to note that we are comparing JTPA services to a mix of no services and other available services, as control group members could still access non-JTPA training. Participants were surveyed by data collection officers between 12 and 36 months after randomization, with an average survey period of 21 months. A second follow-up survey, involving a subset of 5468 participants, focused on the period between the two surveys and was conducted between 23 and 48 months after randomization. Figure 1 in Section G of the Supplementary Material plots a histogram of the observed follow-up time, where a higher censoring rate is indicated by darker shading.

We will focus our attention on the effect of JTPA job training programs on the sample of 3147 single mothers who reported being unemployed at the time of randomization and participated in the initial follow-up interview. The outcome of interest ($T$) is the time between randomization and employment. For participants who only participated in the initial interview, the outcome is fully observed if the individual is employed at the time of the survey $(\Delta_1=1)$. Otherwise, the outcome is censored at the date of the initial interview $(\Delta_2=1)$. For those who participated in the second follow-up interview, the outcome is fully observed if the individual is employed at the time of this second follow-up interview $(\Delta_1=1)$. If not, the outcome is considered to be independently censored at the second interview date $(\Delta_1=\Delta_2=0)$. For a graphical representation of the interview process, see Figure 2 in Section G of the Supplementary Material. Consequently, there may be a dependence between the time until employment and the censoring time if the decision to attend the second follow-up interview is influenced by the individual's employment status between the two interview dates. Note that this means that the individuals who were only invited to the first interview, and were still unemployed by this time, will be censored $(\Delta_1=1)$ instead of administratively censored $(\Delta_1=\Delta_2=0)$. This results from only observing who participated in the second interview, but not who was invited. Because we cannot know which type of censoring occurred, we opted to have them all be classified as censored $(\Delta_1=1)$.

The natural instrumental variable $W$ indicates whether an individual belongs to the control group ($W=0$) or the treatment group ($W=1$). We deem $W$ to be a valid instrument, as it is randomly assigned, moderately correlated with JTPA participation and influences the time to employment only through participation in a JTPA-funded program. The treatment variable $Z$, which is possibly endogenous, indicates whether the individual actually participated in a JTPA program ($Z=0$ for no participation and $Z=1$ for participation). This treatment variable can be confounded due to individuals moving themselves between the treatment and control groups in a non-random way. The covariates considered include the participant’s age (standardized), educational attainment (high school diploma or GED) and race (categorized as white or non-white). Approximately 31\% of the total sample was assigned to the control group. The mean age of this subgroup of single mothers is approximately 28 years old, with 52\% holding a GED or high school diploma and 47\% identifying as white. Notably, 13\% of the women in the control group managed to participate in JTPA services, in contrast to 3\% of the entire control group. From the treatment group, around 70\% participated in JTPA services. Additionally, the mean time to employment appears to be about 43 days shorter for individuals who participated in JTPA training. The censoring and independent censoring rates are comparable between both the control and treatment groups (20\% censoring and 5\% administrative censoring).

We start by splitting the data into two random samples of equal size. The first sample was used to select the specification for the copula and the distribution of $C$. Out of the possible 21 combinations, we select the model with the highest log-likelihood. We can simply use the log-likelihood since each combination of possible copulas and distributions has the same amount of parameters. The results of this selection procedure can be found in Section G of the Supplementary Material for both the proposed and naive estimator. It can be seen that the Frank copula with a log-logistic censoring distribution results in the highest log-likelihood for both of the estimators. The estimation results for both of these methods can be found in Tables \ref{tableJTPAproposed} and \ref{tableJTPAnaive}. In these tables, we also include the estimation results of the model with the second-highest log-likelihood as a type of sensitivity analysis. Note that the $p$-value associated with the null hypothesis $\alpha = \alpha^*$, is calculated as
$$
\frac{1}{B}\sum_{b=1}^B \mathbbm{1}\{\lvert \hat{\alpha}_{b}-\hat{\alpha} \rvert > \lvert \hat{\alpha} - \alpha^* \rvert \},
$$
where $B$ is the amount of bootstrap resamples and $\hat{\alpha}_{b}$ the bootstrap estimate based on the bootstrapped sample $b \in \{1,\dots,B\}$. The $p$-values for the other parameters can be computed similarly. In Tables \ref{tableJTPAproposed} and \ref{tableJTPAnaive}, we test the null hypothesis that the true parameter equals zero, except for $\nu$ where we test the null hypothesis that $\nu = 1$.

The naive estimator appears to underestimate the effect of JTPA services on time
until employment compared to the proposed estimator. At the 5\% significance
level, only the proposed estimator finds a significant effect of JTPA training in reducing unemployment duration, with the CCHR being 1.23. This finding suggests that the individuals participating in the treatment are likely to have a lower inherent ability to secure employment. For both estimators, variables such as age and having a high school diploma or GED do not significantly affect the duration of unemployment. Notably, being white is associated with a significant reduction in unemployment duration for the proposed estimator. Both the proposed and the naive estimator indicate that there is a strong positive dependence between $T$ and $C$, conditional on the treatment and the measured covariates.

\begin{table}[H]
    \centering
    \caption{Estimation results using the proposed estimator for the two models with the highest log-likelihood. The bootstrap standard error (BSE) is based on 500 bootstrap resamples. }
\begin{tabular}{|c|rcc|rcc|}
 \hline
  & \multicolumn{3}{c|}{Frank -- log-logistic} & \multicolumn{3}{c|}{Gaussian -- log-logistic} \\
 \cline{2-7}
  & Estimate & BSE & $p$-value & Estimate & BSE & $p$-value  \\ 
\hline
  $\alpha$ & 0.204& 0.102&  0.040 &  0.272& 0.088 & 0.012    \\ 
  Age & -0.043& 0.034 & 0.204 & -0.059& 0.035 & 0.108      \\ 
   GED &0.023& 0.080 & 0.796 & 0.087 &0.071 & 0.352     \\ 
    white & 0.267 &0.083 & 0.000 & 0.333 &0.079 & 0.000     \\ 
  $\eta_0$ & 6.159& 0.099 & 0.000 &  6.233& 0.089 & 0.000   \\ 
      $\eta_1$ &-0.104& 0.048 & 0.034 & -0.128& 0.044&  0.010   \\ 
  $\eta_2$ &-0.009& 0.021&  0.662 &-0.007& 0.021 & 0.726    \\ 
  $\eta_3$ & -0.019& 0.042 & 0.658 & -0.042& 0.038 & 0.374   \\ 
   $\eta_4$ & -0.101& 0.043&  0.022 &-0.126& 0.042&  0.004   \\ 
  $\nu$ &0.276 & 0.276& 0.028 & 0.253& 0.033 & 0.000    \\ 
  $\tau$ &0.812& 0.082 & 0.006 &  0.803& 0.067&  0.000   \\ 
 \hline
   \end{tabular}
   \label{tableJTPAproposed}
   \end{table}

\begin{table}[H]
    \centering
    \caption{Estimation results using the naive estimator for the two models with the highest log-likelihood. The bootstrap standard error (BSE) is based on 500 bootstrap resamples. }
\begin{tabular}{|c|rcc|rcc|}
 \hline
  & \multicolumn{3}{c|}{Frank -- log-logistic} & \multicolumn{3}{c|}{Gaussian -- log-logistic} \\
 \cline{2-7}
  & Estimate & BSE & $p$-value & Estimate & BSE & $p$-value  \\ 
\hline
 $\alpha$ & 0.002 &0.078 & 0.990  & 0.077& 0.068 & 0.366     \\ 
  Age & -0.038& 0.029 & 0.192   & -0.051 &0.029 & 0.070     \\ 
   GED &-0.043& 0.075 & 0.752 &0.042& 0.061 & 0.566   \\ 
    white &0.197 &0.076 & 0.064 & 0.269 &0.065 & 0.000     \\ 
  $\eta_0$ & 6.062&0.104 & 0.000  & 6.114 &0.073 & 0.000    \\ 
      $\eta_1$ &-0.033& 0.040&  0.636  & -0.058 &0.037 & 0.172    \\ 
  $\eta_2$ & -0.020 &0.019 & 0.344  & -0.008 &0.018 & 0.682     \\ 
  $\eta_3$ &0.005 &0.039 & 0.948 & -0.023 &0.033 & 0.580    \\ 
   $\eta_4$ & -0.075& 0.041 & 0.216 & -0.122& 0.038 & 0.004      \\ 
  $\nu$ & 0.292& 0.025 & 0.000  &  0.274& 0.029 & 0.000   \\ 
  $\tau$ & 0.776 &0.109 & 0.014 & 0.829 &0.059 & 0.000      \\ 
 \hline
   \end{tabular}
   \label{tableJTPAnaive}
   \end{table}  

\newpage

\bibliography{main}

\end{document}